# Rabi-like splitting and refractive index sensing with hybrid Tamm plasmon-cavity modes

S. Jena, R. B. Tokas, S. Thakur, and D. V. Udupa

Atomic & Molecular Physics Division, Bhabha Atomic Research Centre, Mumbai 400 085, India

\* Corresponding author
*E-mail addresses*: shuvendujena9@gmail.com, shujena@barc.gov.in (S. Jena)

**Abstract:** Rabi-like splitting and self-referenced refractive index sensing in hybrid plasmonic-1D photonic crystal structures have been theoretically demonstrated. The coupling between Tamm plasmon and cavity photon modes are tuned by incorporating a low refractive index spacer layer adjacent to the metallic layer to form their hybrid modes. Anticrossing of the modes observed at different values of spacer layer thickness validates the strong coupling between the two modes and causes Rabi-like splitting with different splitting energy. The modes coupling has been supported by coupled mode theory. Rabi-like splitting energy decreases with increasing number of periods ($N$) and refractive index contrast ($\eta$) of two dielectric materials used to make the 1D photonic crystals, and the observed variation is explained by an analytical model. Angular and polarization dependency of the hybrid modes shows that the polarization splitting of the lower hybrid mode is much stronger than that of the upper hybrid mode. On further investigation, it is seen that one of the hybrid modes remains unchanged while other mode undergoes significant change with varying the cavity medium. This nature of the hybrid modes has been utilized for designing self-referenced refractive index sensors for sensing different analytes. For $\eta=1.333$ and $N=10$ in a hybrid structure, the sensitivity increases from 51 nm/RIU to 201 nm/RIU with increasing cavity thickness from 170 nm to 892 nm. For the fixed cavity thickness of 892 nm, the sensitivity increases from 201 nm/RIU to 259 nm/RIU by increasing η from 1.333 to 1.605. The sensing parameters such as detection accuracy, quality factor, and figure of merit for two different hybrid structures ([$\eta=1.333$, $N=10$] and [$\eta=1.605$, $N=6$]) have been evaluated and compared. The value of resonant reflectivity of one of the hybrid modes changes considerably with varying analyte medium which can be used for refractive index sensing.

# 1. Introduction

Surface wave is one special wave form which has very interesting characteristics in optics, electromagnetics, acoustics, etc. Surface wave must propagate along one particular interface. Different variety of surface waves such as surface plasmon polaritons, Zenneck surface waves, Dyakonov waves, and Tamm states can be generated by changing the material and geometries across the interface region [1]. In recent years, optical Tamm state (OTS) is one such surface state that has attracted wider attentions because of its potential applications in nanophotonic devices. For example, Zaky *et al*. [2] have proposed a OTS based refractive index gas sensor with ultra-high sensitivity and low detection limit suitable for biomedical application. Wang *et al.* [3] have shown narrowband and wavelength tunable thermal emitters using OTS in a hybrid structure useful for chemical analysis and bio-sensing. The OTS remains confined near the interface between two highly reflecting media. The sharp resonance in the measured reflection or transmission spectrum reveals the presence of OTS. If the OTS is confined near the interface between a 1DPC and a metallic layer, then it is called as Tamm plasmon (TP) which was theoretically proposed in 2007 [4] and experimentally observed in 2008 [5] by Kaliteevski *et al*. In contrast to surface plasmon resonance (SPR), the TP mode can be excited by both transverse electric (TE) and transverse magnetic (TM) polarized light irrespective of their angle of incidence. The TP modes have gained wide attention due to their potential applications in optical filters, lasers, sensors and optical switches, heat emitters, and light-emitting devices [6-10]. It can be coupled with other resonant modes such as semiconductor excitons for applications in polaritonic devices [11], magnetic or surface plasmon for enhancing magnetic or electric field [12], microcavity modes for narrowband thermal emission [13], and defect modes for induced transparency [14]. The modes coupling generate "hybrid" modes in the system which is evident as a series of non-overlapping resonances in the reflection or transmission spectra. The resonance energies or wavelengths of the hybrid structures are primarily decided by the geometric parameters. The hybrid modes can exhibit interesting properties like Rabi splitting and can be utilized for refractive index sensing application.

When the resonant modes interact weakly, their coupled mode remains almost same as their bare resonant modes. Once the interaction becomes stronger, their coupled state generates new energy eigenvalues and exhibits entirely different optical properties as compared to their uncoupled or bare mode. The universal sign of strong coupling is the anticrossing of the modes in the dispersion curve when the modes approach each other. The energy gap between the modes at the anticrossing point is known as Rabi splitting, and the energy oscillation between the two modes is named as Rabi oscillation [15]. Generally, the strong localization of the TP mode makes strong coupling with the cavity modes, which is confirmed by Rabi-like splitting of the hybrid modes in analogy to cavity quantum-electrodynamic phenomena of Rabi splitting [16]. Quantum mechanically, Rabi splitting occurs as two peaks (dips) in the transmission (reflection) spectrum resulting from a strong coupling between the transition of the two-levels and the cavity mode, when a quantum emitter (two-level atom or a quantum dot or a quantum well) having a specific excitons resonance is placed inside a micro/nano-cavity [17]. Numerous experiments on Rabi splitting have been performed because of its potential applications in the atom detector and infrared photodetectors [18]. In recent years, Rabi-like splitting is explored in hybrid plasmon-exciton systems and hybrid photonic-plasmonic systems [19] for possible applications in various optoelectronic devices by controlling and tailoring the energy distribution [16]. In photonic-plasmonics systems, there has been significant attention on the hybrid modes formed due to simultaneous excitation of the TP mode and cavity mode. Das *et al.* [20] have recently shown the resonant and non-resonant coupling of microcavity mode and Tamm mode in carbon quantum dots embedded 1D photonic crystal structures both theoretically and experimentally. The resonant modes have been tuned by changing the light polarization and by making wedge-shaped metallic layer across the photonic structure [21]. Generally, the modes split into different linear polarizations termed as TE-TM splitting, which is important for phenomena like *all-optical* spin Hall Effect and anisotropic polarization flux [22]. Therefore, the study of modes coupling for different polarizations of light is of vital importance and can be useful in applications of the enhanced light-matter interactions.

Coupling of resonant modes can be exploited to sharpen the resonances, which can be utilized for sensing and filtering applications. It is well known that the detection accuracy of the TP mode-based sensors is better as compared to that of the SPR based sensors [23]. Moreover, both the TP mode and the cavity mode can be excited for both TE and TM polarization of light in contrary to the SPR modes [24]. Though, the Bloch surface wave (BSW) [25] based sensors excited by either TE or TM light are alternative to the metal-based SPR sensors, but they are less flexible like SPR sensors for system portability and miniaturization because of its complex architecture involving gratings or high-index prisms coated with larger number of dielectric layers or optical fibers [26-28]. Therefore, there have been growing interest in optical modes such as TP mode and cavity mode, and their hybrid modes. Ahmed *et al.* [9] have theoretically reported ultra-high sensitivity of 4784 nm/RIU using coupled Tamm/Fano resonant modes in porous silicon photonic crystal sensor. Shaban *et al.* [29] have experimentally reported sensitivity of 50 nm/ RIU with signal to noise ratio of 0.46 in a multilayer Au/SiO$_2$/1D photonic crystal structure. But, these photonic structures do not exhibit self-reference sensing. Samir *et al.* [23] have theoretically realized a coupled TP polariton hybrid-mode based self-referenced refractive index sensor having variable sensitivity of 65 nm/RIU to 180 nm/RIU in the visible region. Most of the proposed hybrid modes-based sensors have very low thickness of analyte medium below 200 nm with sensitivity close to 200 RIU/nm. Submicron thickness analyte medium is very difficult to realize experimentally, therefore efforts have been made to enlarge the thickness of analyte medium with improved sensitivity in the present work.

Here, we have proposed a multilayer plasmonic-photonic hybrid structure: metal/1DPC/cavity/1DPC, which can exhibit Rabi-like modes splitting and self-referenced refractive index sensing of analytes with desired sensing parameters. The strong coupling of TP and cavity modes with varying parameters of the structure such as number of periods in the 1DPC ($N$), refractive index contrast ($\eta=n_H/n_L$, where $n_H$ and $n_L$ represent high and low refractive index layers), thickness of spacer layer, incident angle, and polarization of light have been thoroughly investigated. The hybrid structure is then optimized for refractive index sensing of analytes placed in the cavity having refractive index in the range of 1.33 to 1.49. Two hybrid structures are investigated in the present study for sensing applications. In one structure, the 1DPCs consists of low refractive index contrast materials ($\eta=1.333$) with $N=10$, while the other structure has 1DPCs with $\eta=1.605$ and $N=6$. The sensing parameters are estimated and compared for different thickness of the cavity medium for both the structures. The proposed hybrid structures would be useful in developing self-referenced optical sensors which can be easily integrated in any micro-systems as compared to the SPR based sensors.

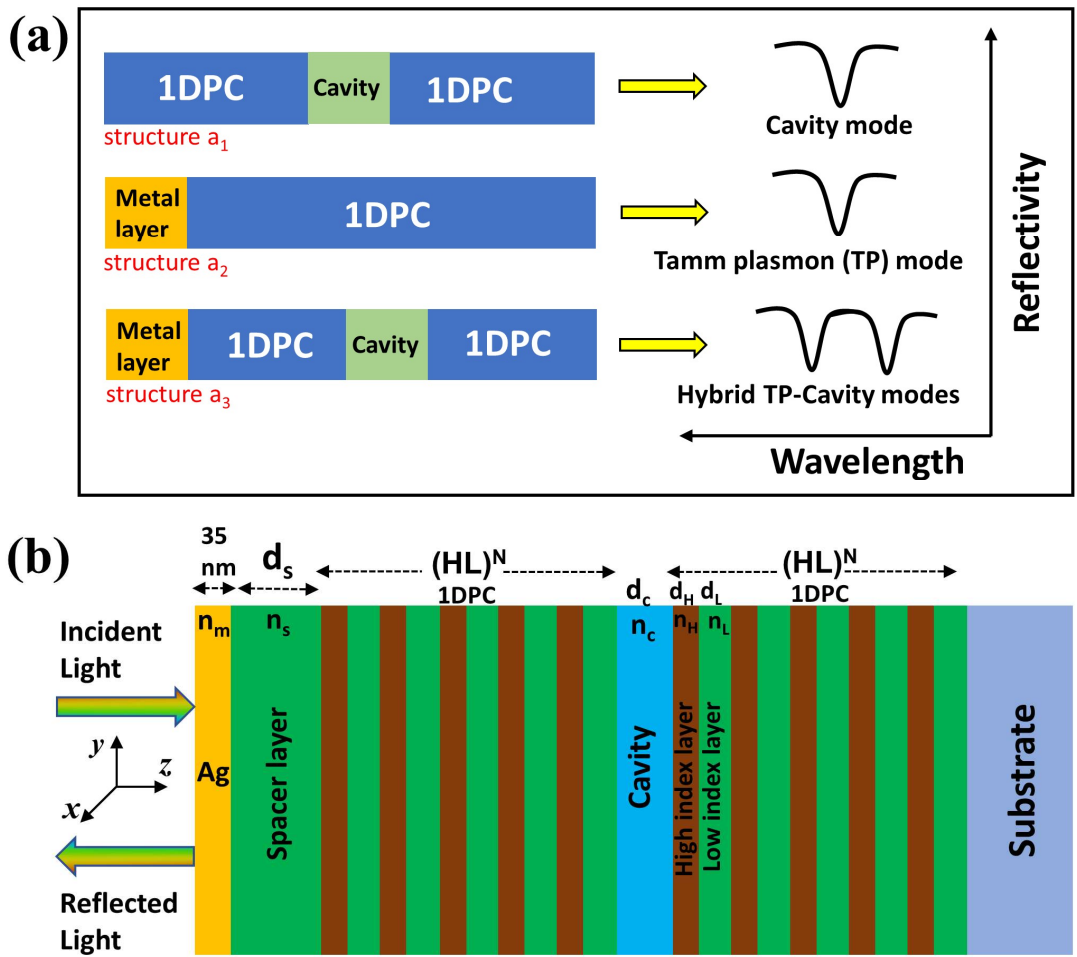

**Fig. 1. (a)** Step by step illustrations towards formation of hybrid Tamm plasmon-cavity modes. The structure $a_1$ forms cavity mode which is observed as a dip in the reflectivity spectrum in the photonic bandgap region. The structure $a_2$ forms Tamm plasmon mode which is observed as a dip in the reflectivity spectrum but with its energy confined at the metal/1DPC interface. The structure $a_3$ forms hybrid Tamm plasmon-cavity modes in which the above two photonic structures $a_1$ and $a_2$ are merged and they appear as a double dip in the reflectivity spectrum. **(b)** Schematic of hybrid photonic structure to investigate coupling of Tamm plasmon and cavity modes, and their utility for refractive index sensing application. The structure consists of a cavity/analyte layer sandwiched between two (HL)$^N$ 1DPCs followed by a spacer layer with Ag metallic layer on top.

## 2. Hybrid Tamm plasmon-cavity modes and theoretical formulation

Coupling of optically resonating modes have been widely investigated in recent years. The most often studied coupled modes include cavity mode and TP mode. It is essential here to briefly discuss the origin of both the modes and their unique features prior to exploring their coupling to form hybrid modes. Fig. 1(a) shows the schematic illustration of each modes and their hybrid modes. Cavity mode is formed between a pair of 1DPCs (structure $a_1$). TP mode is formed at the interface between a thin metallic layer and a 1DPC (structure $a_2$). Both modes are detected as a narrow resonance in the reflection spectrum but with different spatial electric field distribution. The optical energy of the cavity mode is confined within the cavity layer, where as it is localized at the metal/1DPC interface for the TP mode. The sharper the resonance in the spectrum, stronger the field confinement is. When these two individual structures $a_1$ and $a_2$ are merged, the modes are coupled to generate hybrid TP-cavity modes, which are observed as two resonances in the reflection spectrum separated by a finite energy gap (structure $a_3$). In case of hybrid modes, the strong localization of the electric field intensity occurs both at the metal/1DPC interface and in the cavity layer. Fig. 1(b) shows the schematic of the hybrid photonic structure (Ag/Spacer layer/1DPC/Cavity/DPC), which is explored to investigate Rabi-like splitting of modes and refractive index sensing. The 1DPC consists of periodic layers of high index (H) and low index (L) dielectric materials whose refractive index and thickness values are denoted as ($n_H$, $n_L$) and ($d_H$, $d_L$), respectively. Spacer layer is a low index layer (L), and the cavity layer parameters ($n_c$, $d_c$) are chosen as per the requirement. The cavity layer is placed between two 1DPCs to form microcavity structure.

Modes coupling can be clearly understood from the sharp resonances in the optical reflection or transmission spectrum of the hybrid structure, which can be computed using transfer matrix method (TMM) [30-32]. A multilayer consists of $m$ layers with $m+1$ interfaces. In TMM, the right and left electric field components of a layer in the multilayer is related by a 2x2 matrix utilizing the fact that the electric field wave equations are linear, and the tangential component of the electric field is continuous [30]. Each layer $j$ ($j$ = 1, 2, 3,….,$m$) has a thickness $d_j$ and its complex refractive index is expressed as $\tilde{n}_j = n_j + ik_j$ which is a function of wavelength of light. The electric field components at ambient side ($j$ = 0) and substrate side ($j$ = $m+1$) are inter-related by the transfer matrix $S$ as follows:

$$\begin{bmatrix} E_0^+ \\ E_0^- \end{bmatrix} = S \begin{bmatrix} E_{m+1}^+ \\ E_{m+1}^- \end{bmatrix} \quad (1)$$

where $E_0^\pm$ and $E_m^\pm$ are the electric field components travelling in the positive (+) and negative (-) direction at $j$ = 0 and $j$ = $m+1$, respectively. The transfer matrix ($S$) is obtained by multiplying refractive matrix $I_{j-1,j}$ of each interface and phase matrix $L_j$ of each layer as follows [30]:

$$S = \begin{bmatrix} S_{11} & S_{12} \\ S_{21} & S_{22} \end{bmatrix} = \left( \prod_{j=1}^{m} I_{j-1,j} L_j \right) I_{m,m+1} \quad (2)$$

The refractive matrix or interface matrix between $(j-1)^{th}$ and $j^{th}$ layers that describes refraction at $j^{th}$ layer interface is given as follows

$$I_{j-1,j} = \frac{1}{t_{j-1,j}} \begin{bmatrix} 1 & r_{j-1,j} \\ r_{j-1,j} & 1 \end{bmatrix} \quad (3)$$

The phase matrix or layer matrix that describes propagation of electric field wave through $q^{th}$ layer is given as follows

$$L_j = \begin{bmatrix} e^{-i\delta_j d_j} & 1 \\ 1 & e^{i\delta_j d_j} \end{bmatrix} \quad (4)$$

Here $t_{j-1,j} = 2y_{j-1}/(y_{j-1}+y_j)$ and $r_{j-1,j} = (y_{j-1}-y_j)/(y_{j-1}+y_j)$ are the complex Fresnel transmission and reflection co-efficients at $(j-1, j)$ interface, respectively, $y_j = \tilde{n}_j \cos\theta_j$ for S-polarized or TE waves, $y_j = \tilde{n}_j/\cos\theta_j$ for P-polarized or TM waves, $\delta_j = 2\pi\tilde{n}_j \cos\theta_j/\lambda$ is the phase change experienced by the wave in travelling $j^{th}$ layer, $\cos\theta_j = \sqrt{1-(n_0 \sin\theta_0/\tilde{n}_j)^2}$, $n_0 \approx 1$ is the refractive index of the incident medium (air), $\theta_0$ is the angle of incidence, and $\theta_j$ is the angle of refraction in $j^{th}$ layer [33]. The reflection and transmission co-efficients are given by $r = E_0^-/E_0^+ = S_{21}/S_{11}$ and $t = E_{m+1}^+/E_0^+ = 1/S_{11}$, respectively. Transmissivity and reflectivity of the multilayer structure are expressed as $T_M = |t|^2 \operatorname{Re}(\tilde{n}_{m+1})/n_0$ and $R_M = |r|^2$, respectively. Typical thickness of glass substrates is in the range 0.5-10 mm $\gg \lambda$. Therefore, the reflection and transmission at the air-substrate and substrate-multilayer interfaces must be considered; consequently total transmission and total reflection of the multilayer/substrate system can be expressed as follows [34]:

$$T = \frac{T_S T_M e^{-2\beta_s}}{1 - R_S R_M e^{-4\beta_s}} \quad (5)$$

$$R = R_M + \frac{R_S T_M^2 e^{-4\beta_s}}{1 - R_S R_M e^{-4\beta_s}} \quad (6)$$

where $T_S = |4n_0\tilde{n}_{m+1}/(n_0+\tilde{n}_{m+1})^2|$ and $R_S = |(n_0-\tilde{n}_{m+1})^2/(n_0+\tilde{n}_{m+1})^2|$ are transmission and reflection, respectively at substrate-air interface, $\tilde{n}_{m+1} = \tilde{n}_s = n_s + ik_s$ is the complex refractive index of the substrate, and $\beta_s = 2\pi k_s d_s/\lambda$ is the phase thickness related to absorption in the substrate. $k_s$ and $d_s$ are the extinction co-efficient and thickness of the substrate, respectively.

The electric field within any arbitrary layer $q$ is computed by considering the division of the multilayer structure into two subsections separated by the layer $q$, as a result the total system transfer matrix can be expressed as [35]

$$S = S_q' L_q S_q'' \quad (7)$$

with $S_q' = \begin{bmatrix} S_{q11}' & S_{q12}' \\ S_{q21}' & S_{q22}' \end{bmatrix} = \left( \prod_{j=1}^{q-1} I_{j-1,j} L_j \right) I_{q-1,q}$ (8)

and $S_q'' = \begin{bmatrix} S_{q11}'' & S_{q12}'' \\ S_{q21}'' & S_{q22}'' \end{bmatrix} = \left( \prod_{j=q+1}^{m} I_{j-1,j} L_j \right) I_{m,m+1}$ (9)

The complex reflection and transmission co-efficients of layer $q$ in terms of matrix elements can be defined as $r_q' = S_{q21}'/S_{q11}'$, $r_{q-}' = -S_{q12}'/S_{q11}'$, $t_q' = 1/S_{q11}'$, $r_q'' = S_{q21}''/S_{q11}''$ and $t_q'' = 1/S_{q11}''$. The electric field travelling in the forward (positive) and backward (negative) direction in the layer $q$ at the left interface ($q$-1, $q$) is related to the incident plane wave $E_0^+$ as follows [30]:

$$t_q^+ = \frac{E_q^+}{E_0^+} = \frac{t_q'}{1 - r_{q-}' r_q'' e^{i 2 \delta_q d_q}}$$ (10)

$$t_q^- = \frac{E_q^-}{E_0^+} = \frac{t_q' r_q'' e^{i 2 \delta_q d_q}}{1 - r_{q-}' r_q'' e^{i 2 \delta_q d_q}} = t_q^+ r_q'' e^{i 2 \delta_q d_q}$$ (11)

Using equation (10) and (11), the depth (z) dependent electric field distribution in any arbitrary position in $q^{th}$ layer can be expressed in terms of incident plane wave $E_0^+$ as follows

$$E_q(z) = E_q^+(z) + E_q^-(z) = [t_q^+ e^{i \delta_q z} + t_q^- e^{-i \delta_q z}] E_0^+$$ (12)

The normalized electric field intensity is estimated using Eq. (12). The spatial localization of the electric fields in the multilayer determines the existence of optical Tamm mode or cavity mode or their coupled modes. Electric field distribution is therefore an essential parameter that needs to be determined for analyzing the modes present in the hybrid structure.

## 3. Results and discussion

The plasmonic-photonic crystal heterostructure as shown in Fig. 1(b) is systematically analyzed to realize Rabi-like splitting and refractive index sensing using the hybrid modes, and are discussed below in detail. Rabi-like splitting is expressed conveniently using reflectivity as a function of energy of the light, whereas it is convenient to use reflectivity as a function of wavelength for refractive index sensing application. One can always be converted to other through the relation *Energy*=$\hbar\omega$=$hc/\lambda$, where $\omega$ and $\lambda$ are the angular frequency and wavelength of light, respectively.

### 3.1. Mode coupling

The proposed hybrid photonic structure (schematic shown in Fig. 1(b)) for tunable mode-coupling consists of thin Ag layer followed by low index spacing layer covering a cavity structure made of a low index layer sandwiched between two 1DPCs placed on a transparent glass substrate. The 1DPC structure comprises $N$=5 periods of alternative high index (H)/low index (L) layers. The hybrid heterostructure is: Ag/S/(HL)$^5$/C/(HL)$^5$/Substrate. The refractive index of both S and C layers is same i.e. $n_c$=$n_s$=1.47. The thickness values of Ag layer, S layer, and C layer are $d_m$ = 35 nm, $d_s$=141 nm, and $d_c$ = 85 nm, respectively. Both the H and L layers are assumed as lossless and dispersion free dielectrics with refractive index and thickness values of ($n_H$=1.96, $d_H$=64 nm) and ($n_L$=1.47, $d_L$=85 nm), respectively. The complex permittivity of Ag as a function of frequency is determined using Drude model as expressed in equation (21).

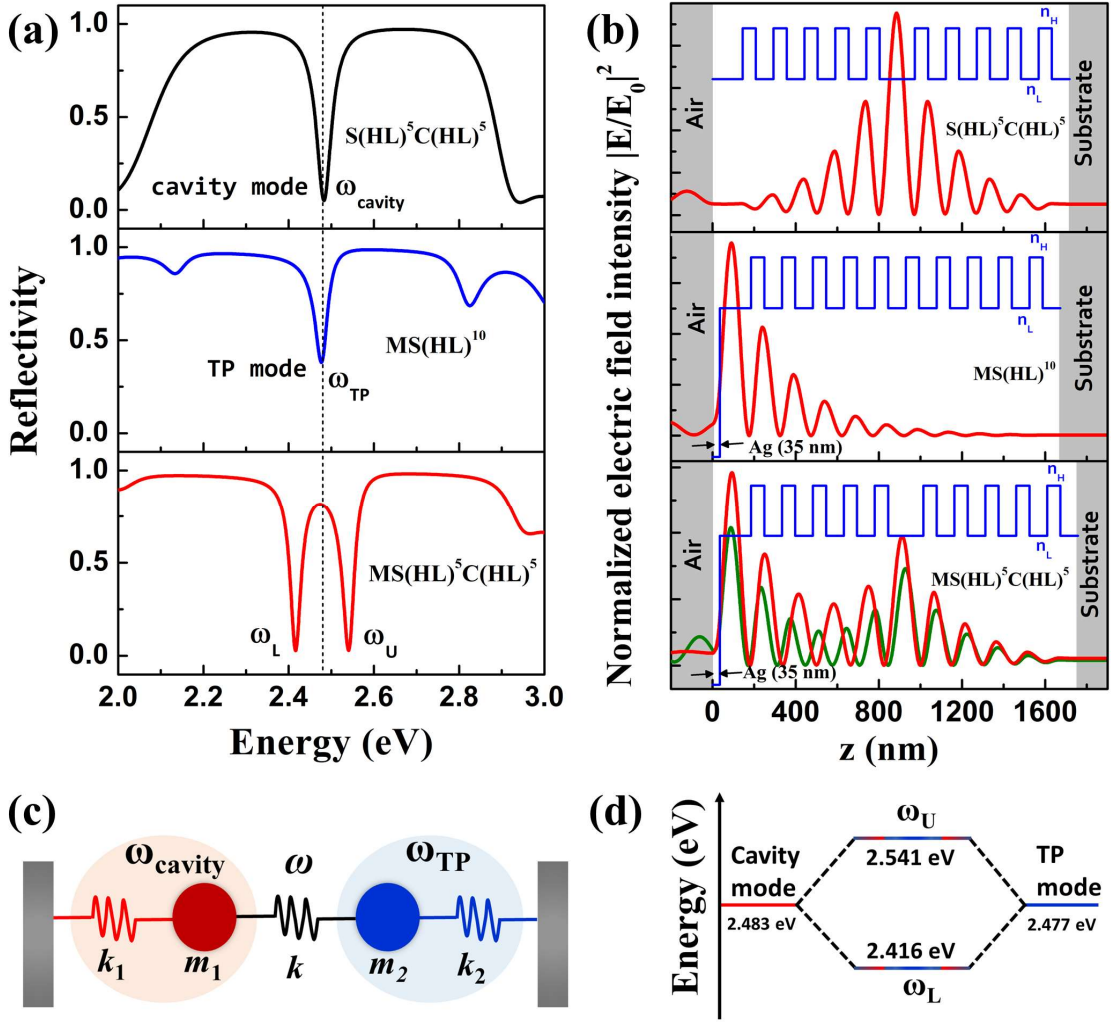

**Fig. 2.** Proposed hybrid structure is MS(HL)$^5$C(HL)$^5$, where M is Ag thin layer having thickness $d_m$=35 nm, H is a high index ($n_H$=1.96) thin layer having thickness $d_H$=64 nm, L is a low index ($n_L$=1.47) thin layer having thickness $d_L$=85 nm, S is a low index ($n_S$=1.47) spacer layer having thickness $d_s$ = 141 nm, and C is a cavity layer having thickness $d_c$=85 nm filled with medium having refractive index $n_c$=1.47. **(a)** Reflection spectra of photonic structure without metallic layer (bare cavity mode), without cavity layer (TP mode), and with both metallic and cavity layers (hybrid TP-cavity modes), respectively. **(b)** The refractive index profile and the normalized electric field intensity distribution in the top plot corresponds to the cavity mode localized between two 1DPCs at light energy of 2.483 eV, the middle plot corresponds to the Tamm-plasmon mode localized at the interface between Ag layer and a 1DPC at light energy of 2.477 eV, and the bottom plot corresponds to the hybrid Tamm plasmon-cavity modes at $\omega_L$=2.416 eV (dotted green line) and $\omega_U$=2.541 eV (solid red line), respectively. Schematic of **(c)** a coupled oscillator model, and **(d)** energy diagram for coupling between TP and cavity modes.

Fig. 2(a) shows the calculated reflection spectra of MS(HL)$^5$C(HL)$^5$ structure without M layer, without C layer, and with both M and C layers, respectively. The photonic bandgap lies in the energy range of 2.2-2.7 eV with center at Bragg frequency $\omega_0 \approx 2.48$ eV. The optimized thickness values of M layer (Ag) and S layer are 35 nm and 141 nm, respectively, in order to obtain desired TP mode at the center of the photonic bandgap. Without spacing layer, the TP mode deviates significantly from the resonant cavity mode (see supplementary material Fig. S1), as a result the coupling between the TP mode and cavity mode becomes poor. Therefore, the spacing layer thickness is very crucial for strong coupling of modes. The TP mode is also very sensitive to M layer thickness (see supplementary material Fig. S2). When thickness of the metallic layer is more than its penetration depth or skin depth, the layer becomes effectively opaque, and consequently the structure does not support the TP mode at the metal/1DPC interface due to insufficient light. The thickness hence needs to be less than the skin depth to get TP mode. The TP mode undergoes red-shift with decreasing M layer thickness, and then disappears when M layer thickness is such that it does not reflect sufficient light to sustain the mode at the interface. Therefore, the optimized value of M layer thickness required for sustaining TP mode is 35 nm in the present structure. Similarly, the thickness of S layer is also equally important for sustaining the TP mode at the desired energy or wavelength. The minimum thickness required to make TP mode very close to the Bragg frequency is $d_s$=141 nm, which is explained later in detail. The resonance generates dips at 2.483 eV and 2.477 eV in the reflection spectrum for cavity mode without M layer and TP mode without C layer, respectively, close to the center of the bandgap. The reflectivity is around 1% with C layer, while it is around 40% with M layer around 2.48 eV. The high reflectivity of the Ag metallic layer makes the dip smaller in case of TP mode. In the case of MS(HL)$^5$C(HL)$^5$ structure with both M and C layers, one would expect that the two modes will be merged to a give a sharper resonant reflectivity at $\omega_0 \approx$ 2.48 eV. However, this does not happen since the modes get strongly coupled in such a way that there is no resonance reflection at $\omega_0 \approx$ 2.48 eV. Instead of a single resonant mode, two resonant hybrid modes are generated at 2.416 eV and 2.541 eV, respectively symmetric around 2.48 eV. The properties of hybrid TP-cavity modes are entirely different from their individual uncoupled modes. The spatial confinement of such modes can be established by numerically estimating electric field intensity. Fig. 2(b) shows the electric field intensity distribution for the S(HL)$^5$C(HL)$^5$ structure (cavity mode at 2.483 eV), MS(HL)$^{10}$ structure (TP mode at 2.477 eV), and MS(HL)$^5$C(HL)$^5$ structure (hybrid TP-cavity modes at 2.541 eV and 2.416 eV). The electric field intensity at cavity mode of 2.483 eV is found spatially localized in the cavity C layer of the S(HL)$^5$C(HL)$^5$ structure, confirms the presence of cavity mode formed between the two (HL)$^5$ 1DPCs. The electric field intensity at TP mode of 2.477 eV is found spatially localized in the S layer of MS(HL)$^5$C(HL)$^5$ structure, confirms the presence of TP mode formed between the M layer and the (HL)$^5$ 1DPC. This shows that TP mode is a surface mode that propagates along the surface unlike the cavity mode where light is confined within the cavity layer of the structure. The electric field intensity at the two resonant hybrid modes of 2.541 eV and 2.416 eV is found localized both at the metal/1DPC interface, and in the cavity layer of the MS(HL)$^5$C(HL)$^5$ structure. The M layer thickness dependent reflectivity calculation indicates (see supplementary material Fig. S2) that, the energy $\omega_U$=2.541 eV corresponds to the cavity mode part whereas the energy $\omega_L$=2.416 eV corresponds to TP mode part of the hybrid TP-cavity modes. The value of electric field intensity localized in the S layer as well as C layer for 2.541 eV is higher as compared to that for 2.416 eV. The field confinement both in the S and C layer clearly indicates that the TP and cavity modes are strongly coupled in the exact crossing region. The hybrid TP-cavity modes can be classically represented by two coupled oscillators having uncoupled energy eigenvalues $\omega_{cavity}$ and $\omega_{TP}$, respectively as shown in the Fig. 2(c). The strong interaction between bare cavity and bare TP modes generate hybrid mode energy levels ($\omega_U$ and $\omega_L$) as shown in Fig. 2 (d) that are indistinguishable unlike in their uncoupled mode, and exhibits entirely new features.

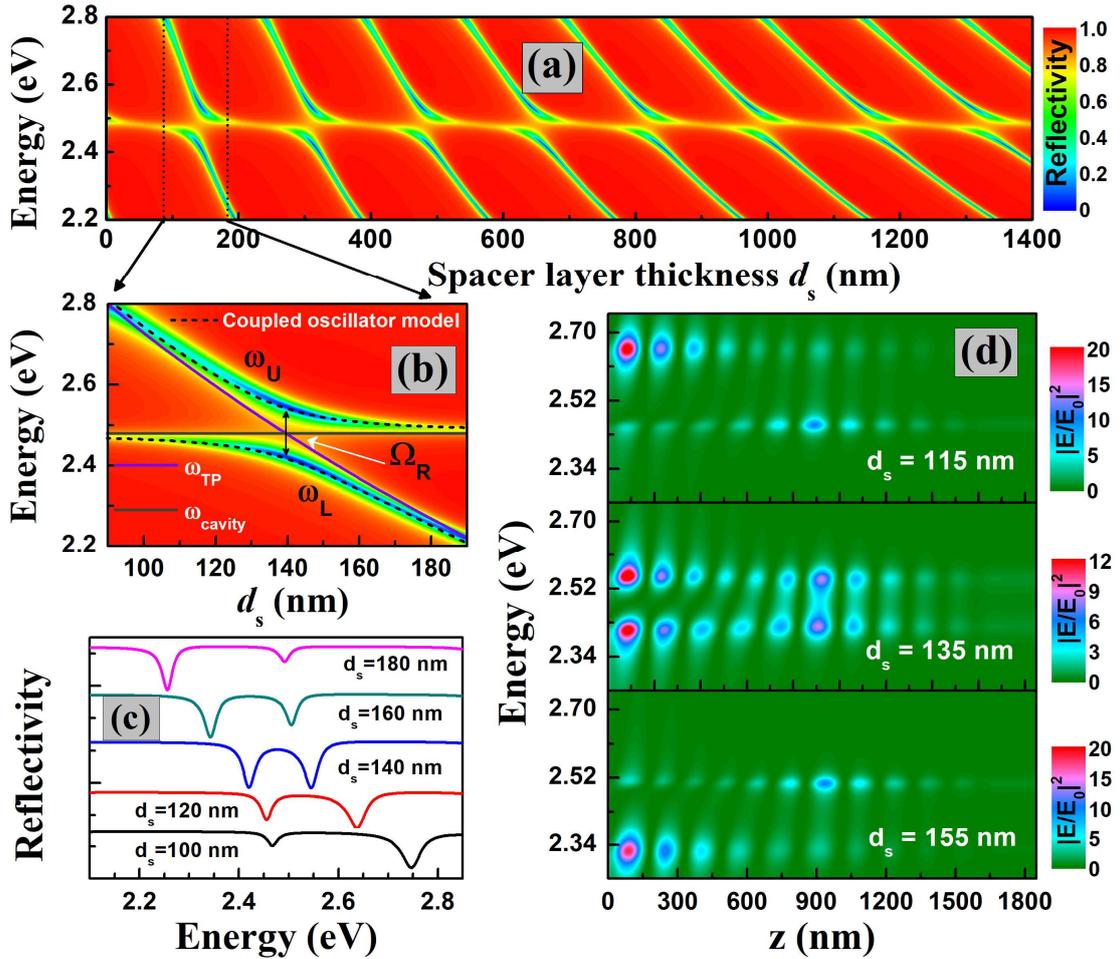

**Fig. 3.** Reflectivity contour of the hybrid structure as a function of incident light energy and S layer thickness $d_s$ **(a)** over a wider range with several anticrossing, and **(b)** over a limited range of 90 nm-190 nm with only one anti-crossing. The blue color in the contour plot represents the positions of the reflectivity dip of the hybrid modes as a function of $d_s$ obtained using TMM and are fitted using coupled oscillator model (dotted black lines). The bare TP mode (solid navy-blue line) and the bare cavity mode (solid grey line) are also presented. **(c)** Reflectivity spectra for different values of $d_s$. **(d)** Normalized electric field intensity $|E/E_0|^2$ (arb. unit) as a function of energy and depth (z) in the structure for three different values of $d_s$ = 115 nm, 135 nm, and 155 nm around the coupling region.

The coupling between the TP and cavity modes has been tuned by varying thickness of spacing layer ($d_s$). The reflectivity as a function of energy and $d_s$ of the hybrid structure is plotted in contour Fig. 3 (a). It shows that there are several anticrossing of the two modes at different values of $d_s$ resulting to different values of Rabi-like splitting of the modes, which is explained later in detail. The behaviour of coupled TP-cavity modes around a particular spacer layer thickness of $d_s$=141 nm is thoroughly investigated using coupled oscillator model as illustrated in Fig. 3 (b). The figure shows the variation of bare TP mode ($\omega_{TP}$), bare cavity mode ($\omega_{cavity}$), and their coupled hybrid modes ($\omega_U$ and $\omega_L$) as a function of $d_s$. The bare TP mode exhibits redshift with increasing $d_s$, while the bare cavity mode does not change with $d_s$ and remains fixed at ~2.486 eV. However, the behavior of coupled modes in hybrid photonic structure is very interesting. The hybrid modes deviate from their bare TP and cavity modes in the strong coupling region *i.e.* 130 nm ≤ $d_s$ ≤ 150 nm, while their variation with $d_s$ is similar to that of the bare modes in either side of the strong coupling region. The TP mode and the cavity mode appear to strongly repel each other away from their nearly common bare mode energy of ~ 2.48 eV for $d_s$=141 nm, which results in the anticrossing of the two modes as they come closer. This anticrossing is the indication of strong coupling. The TP-cavity hybrid modes $\omega_U$ and $\omega_L$ are located at the energy of 2.416 eV and 2.541 eV, respectively. The separation between the two modes $\omega_U$ and $\omega_L$ exactly at the anticrossing region is 125 meV, known as Rabi-like splitting. The splitting energy is considerably greater than their bare mode linewidth, which further confirms the occurrence of strong coupling between TP and cavity modes [36]. Hence, it can be seen that the interaction between TP mode and cavity mode in a hybrid photonic structure can be controlled by varying S layer thickness.

The mode coupling in the hybrid structure can be better understood by considering coupled oscillator model. The TP mode and the cavity mode can be considered as two oscillators coupled in the hybrid structure. The energy eigenvalues ($\omega$) of the coupled oscillators can be obtained by solving the following equation [37]:

$$\begin{vmatrix} \omega - \omega_{cavity} & \Omega_{cT} \\ \Omega_{cT} & \omega - \omega_{TP} \end{vmatrix} = 0 \qquad (13)$$

where $\omega_{cavity}$ and $\omega_{TP}$ are the eigen energies of the bare cavity mode and bare TP mode, respectively. $\Omega_{cT}$ is the coupling strength between the modes. From equation (13), we can get

$$(\omega - \omega_{cavity})(\omega - \omega_{TP}) = \Omega_{cT}^2 \qquad (14)$$

The spacing layer thickness $d_s$ adjacent to the Ag metallic layer is adjusted to tune the TP mode. Therefore, $\omega_{TP}$ is considered as a function of $d_s$ *i.e.* $\omega_{TP}(d_s)$. Hence, the two solutions of the quadratic equation (14) can be expressed as follows:

$$\omega_U(d_s) = \frac{1}{2}\left[(\omega_{cavity} + \omega_{TP}(d_s)) + \sqrt{(\omega_{cavity} - \omega_{TP}(d_s))^2 + 4\Omega_{cT}^2}\right] \qquad (15)$$

$$\omega_L(d_s) = \frac{1}{2}\left[(\omega_{cavity} + \omega_{TP}(d_s)) - \sqrt{(\omega_{cavity} - \omega_{TP}(d_s))^2 + 4\Omega_{cT}^2}\right] \qquad (16)$$

The upper and lower hybrid modes have been theoretically fitted using equation (15) and (16), respectively as shown in Fig. 3 (b). This shows that the resonant energies of the hybrid modes obtained using coupled oscillators model are in excellent agreement with that of the values obtained from transfer matrix method. The upper hybrid mode $\omega_U$ behaves like TP mode for lower $d_s$ values, and gradually changes to behave like cavity mode with increasing $d_s$ value. The lower hybrid mode $\omega_L$ behaves like cavity mode for lower $d_s$ values, and gradually changes to behave like TP mode for increasing $d_s$ value. The reflectivity spectrum for different $d_s$ values are plotted in Fig. 3(c) to understand the modes coupling in a system. one can see that for spacer thickness values of $d_s$=180 nm, 160 nm, 120 nm, and 100 nm, the coupled modes are formed by an admixture of unequal weight of TP mode and cavity mode, and the width of the two reflectivity dips are not same. This is a signature of weak coupling system. But in case of $d_s$=140 nm, the couple modes have equal weight with identical amplitude and linewidth of the reflectivity dip indicating the strong coupling behaviour. This implies that a system can be made either weakly coupled or strongly coupled by tuning $d_s$ value. Hence, it can be said that the spacer layer acts as a modulator for tailoring the modes of the coupled system. The mechanism behind this can be better understood by analyzing the electric field intensity in the structure. To do this, we have chosen three values of $d_s$: 115 nm, 135 nm, and 155 nm. The thickness values of 115 nm and 155 nm are lower and higher, respectively than the strong coupling thickness region, whereas the thickness of 135 nm is in the strong coupling region. In order to compare the effect of $d_s$, electric field distribution as a function of energy and depth (z) for the three $d_s$ values are plotted in Fig. 3(d). For $d_s$=115 nm, the modes are uncoupled, and the cavity mode is completely suppressed for the uncoupled upper energy mode, while the TP mode is completely absent in case of the uncoupled lower energy mode. The observations are exactly opposite for $d_s$=155 nm. The modes are strongly coupled for $d_s$=135 nm, and the amplitude of TP mode is larger than that of the cavity mode for both the lower as well as upper energy modes. This shows that there exists a competing phenomenon between the two modes in their coupled states.

It is seen that the bare cavity mode does not change much with varying spacer layer thickness ($d_s$) over a large range, while the bare TP mode shifts significantly with $d_s$ (see supplementary material Fig. S3). The coupling between the TP mode and cavity mode is primarily due to the tuning of the TP mode with varying $d_s$. Therefore, the anti-crossing indicating strong coupling observed at different values of $d_s$ in Fig. 3 (a) can be explained only investigating the TP mode behaviour. In the MS(HL)$^N$, the light trapped in the spacer layer placed between two mirrors like media (metal layer M and (HL)$^N$ 1DPC) gets resonated to form TP mode. The reflection and transmission co-efficients of the Fabry-Perot cavity like M/S/1DPC structure as shown in Fig. 4 are given by [38, 39]

$$r = r_{0M} + \frac{t_{0M} t_{SM} r_{SP} e^{2i\varphi_s}}{1 - r_{SM} r_{SP} e^{2i\varphi_s}} \qquad (17)$$

$$t = \frac{t_{SM} t_{SP} e^{2i\varphi_s}}{1 - r_{SM} r_{SP} e^{2i\varphi_s}} \qquad (18)$$

where $r_{0M}$ and $t_{0M}$ are the reflection and transmission co-efficients of air/metal layer interface, $r_{SM}$ and $t_{SM}$ are the reflection and transmission co-efficients of metal/spacer layer interface, $\varphi_s = n_s d_s \omega/c$ is phase shift of the light passing through the spacer layer, and $r_{SP}$ is the reflection co-efficient of spacer layer/ (HL)$^N$ 1DPC interface. The strong coupling between TP and cavity modes occurs near $\omega_0$ at specific $d_s$ values where the TP mode is tuned to $\omega_0$, which is detected as a dip (peak) in reflection (transmission) spectrum. As per the equation (17) and (18), the following condition must be satisfied at those $d_s$ values:

$$r_{SM} r_{SP} e^{2i\varphi_s} = 1 \qquad (19)$$

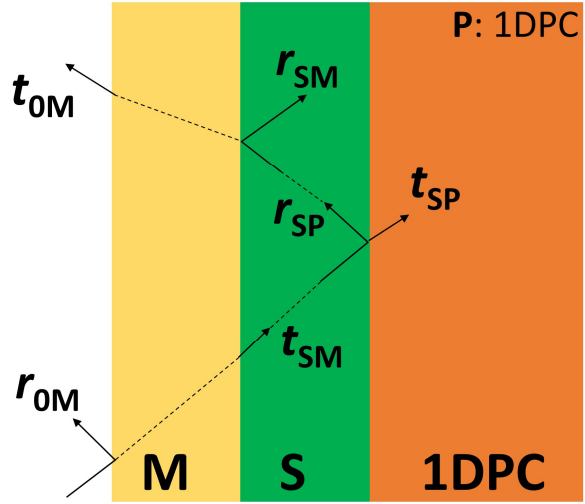

**Fig. 4:** Illustration of Fresnel reflection and transmission co-efficients at the interfaces of Fabry-Perot cavity like M/S/1DPC structure resonating the TP mode.

The reflection co-efficient $r_{SM}$ is given by

$$r_{SM} = \frac{n_s - n_m}{n_s + n_m} \qquad (20)$$

where $n_s$=1.47 is the refractive index of spacer layer, and $n_m$ is the complex refractive index of silver layer given by [40]

$$n_m^2 = \varepsilon_{Ag} = \varepsilon_\infty - \frac{\omega_p^2}{\omega^2 - i\omega\gamma} \qquad (21)$$

where $\varepsilon_\infty$ = 5 is the infinite-frequency dielectric constant, $\omega_p$=1.36x10$^{16}$ rad/s (~8.95 eV) is the plasma frequency, and $\gamma$=6.6x10$^{13}$ rad/s is the effective electron scattering rate. Since the plasma frequency $\omega_p$ of the metal is much larger than the frequency of light considered in the present case (visible light) *i.e.* $\omega \ll \omega_p$, and scattering rate ($\gamma$) is relatively small, therefore equation (21) can be approximated as

$$n_m \approx i\sqrt{x}\,\frac{\omega_p}{\omega} \qquad (22)$$

where $x = 1 - \left(\varepsilon_\infty \frac{\omega^2}{\omega_p^2}\right) \qquad (23)$

Putting equation (22) in equation (20), we will get

$$r_{SM} = -\frac{(i n_s \omega / \sqrt{x}\,\omega_p) + 1}{(i n_s \omega / \sqrt{x}\,\omega_p) - 1}$$

$$r_{SM} \approx -\left(\frac{in_s\omega}{\sqrt{x}\omega_p}+1\right)\left(\frac{in_s\omega}{\sqrt{x}\omega_p}+1\right) \approx -\left(1+\frac{2in_s\omega}{\sqrt{x}\omega_p}\right) \qquad (24)$$

Adding higher order terms of ($\omega/\omega_p$<1) in equation (24), we will get

$$r_{SM} \approx -\left(1+\frac{2in_s\omega}{\sqrt{x}\omega_p}+\frac{1}{2!}\left(\frac{2in_s\omega}{\sqrt{x}\omega_p}\right)^2+\frac{1}{3!}\left(\frac{2in_s\omega}{\sqrt{x}\omega_p}\right)^3+............\right)$$

$$r_{SM} \approx -\exp\left(\frac{2in_s\omega}{\sqrt{x}\omega_p}\right) \approx \exp\left[i\left(\pi+\frac{2n_s\omega}{\sqrt{x}\omega_p}\right)\right] \qquad (25)$$

Considering $N$ is large number of periods in the 1DPC, the reflection co-efficient of the wave with frequency $\omega \approx \omega_0$ at the spacer layer/1DPC interface is given by [4]

$$r_{SP} \approx -\exp\left(i\beta\frac{\omega-\omega_0}{\omega_0}\right) = \exp\left[i\left(\pi+\beta\frac{\omega-\omega_0}{\omega_0}\right)\right] \qquad (26)$$

where $\beta=\pi/(\eta-1)$, and $\eta=n_H/n_L$ is the refractive index contrast in the 1DPC. Equation (19) can be written as

$$\pi+\frac{2n_s\omega}{\sqrt{x}\omega_p}+\pi+\beta\frac{\omega-\omega_0}{\omega_0}+\frac{2n_s d_s\omega}{c}=m\pi \qquad (27)$$

where $m$=0, 2, 4, 6, 8, 10, …. even integers. As the TP mode approaches to the Bragg frequency i.e. $\omega=\omega_{TP}\approx\omega_0$ in the anti-crossing regime, equation (27) can be now expressed as

$$\pi+\frac{2n_s\omega_0}{\sqrt{x_0}\omega_p}+\frac{2n_s d_s\omega_0}{c}+\pi=m\pi \qquad (28)$$

Using $\varepsilon_\infty$=5, $\omega_0$=2.48, and $\omega_p$=8.95, the value of $x_0 = 1-\left(\varepsilon_\infty\frac{\omega_0^2}{\omega_p^2}\right) \approx 0.6$.

Equation (28) can be further reduced to

$$d_s = \frac{(m-2)\pi c}{2n_s\omega_0} - \frac{c}{\sqrt{x_0}\omega_p} \qquad (29)$$

As per the equation (29), $d_s$≈141 nm, 311 nm, 481 nm, 651 nm, 821 nm, 991 nm,…........., respectively for $m$=4, 6, 8, 10, 12, 14,……….., which exactly matches to the anti-crossing $d_s$ values observed in Fig. 3 (a). This shows that the thickness of spacer layer adjacent to the silver layer is chosen to tune the TP mode to $\omega_0$ in order to obtain strong coupling with the cavity mode. At these specific $d_s$ values, the phase shift of the light passing through the spacer layers helps in exchanging the energy between the TP mode and the cavity mode. The energy splitting between the coupled TP-cavity modes is larger than the sum of their line widths. It means the exchange energy rate is more than their individual mode losses, therefore the strong coupling between the modes sustains at these $d_s$ values. Therefore, the $d_s$ value as per the equation (29) can be chosen for depositing the spacing layer in order to achieve strong coupling between the modes in the fabricated hybrid structure.

## 3.2. Rabi-like splitting

The TP and cavity modes are allowed to couple around a common energy $\omega_0$ by choosing a spacer layer thickness $d_s$=141 nm resulting to Rabi-like splitting in the strong coupling region where the gap between upper and lower mode becomes minimum. This minimum gap energy is known as Rabi energy ($\Omega_R$). It is seen that Rabi energy varies with change in number of periods ($N$) and refractive index contrast between the dielectric layers ($\eta$) in the 1DPC. The variation of hybrid TP-cavity modes with number of periods (N) in the (HL)$^N$ 1DPC for a given refractive index contrast of $\eta = n_H/n_L = 1.96/1.47 = 1.33$ has been numerically estimated and plotted in Fig. 5(a). The figure shows that the linewidth of the resonant coupled modes becomes narrower with increasing $N$ values. The resonant reflectivity value increases with increasing $N$, but the value remains same for both the hybrid modes. The value of $\Omega_R$ decreases dramatically with increasing $N$ values, and becomes almost constant for higher value of $N$ as shown in Fig. 5(c). The reflectivity of the hybrid structure with increasing value of $\eta$ for a fixed value of $N=5$ has been calculated and plotted in Fig. 5(b). The trend of variation of $\Omega_R$ in hybrid modes with $\eta$ is almost identical to that of $N$ as shown in Fig. 5(d). The variation of Rabi-like splitting with $N$ and $\eta$ has been explained using an analytical model proposed by Kaliteevski *et al.* [41] based on transfer matrix elements for complex amplitudes of forward and backward propagating waves. They have shown that the transfer matrix equation for the hybrid structure can be approximated as linear function of $\omega$ and can be reorganized in the form of $(\omega - \omega_{cavity})(\omega - \omega_{TP}) = \Omega_{cT}^2$, characteristic of two coupled oscillators. For $\omega_{TP} \approx \omega_0$, the value of $\Omega_{cT}$ is given by [41]

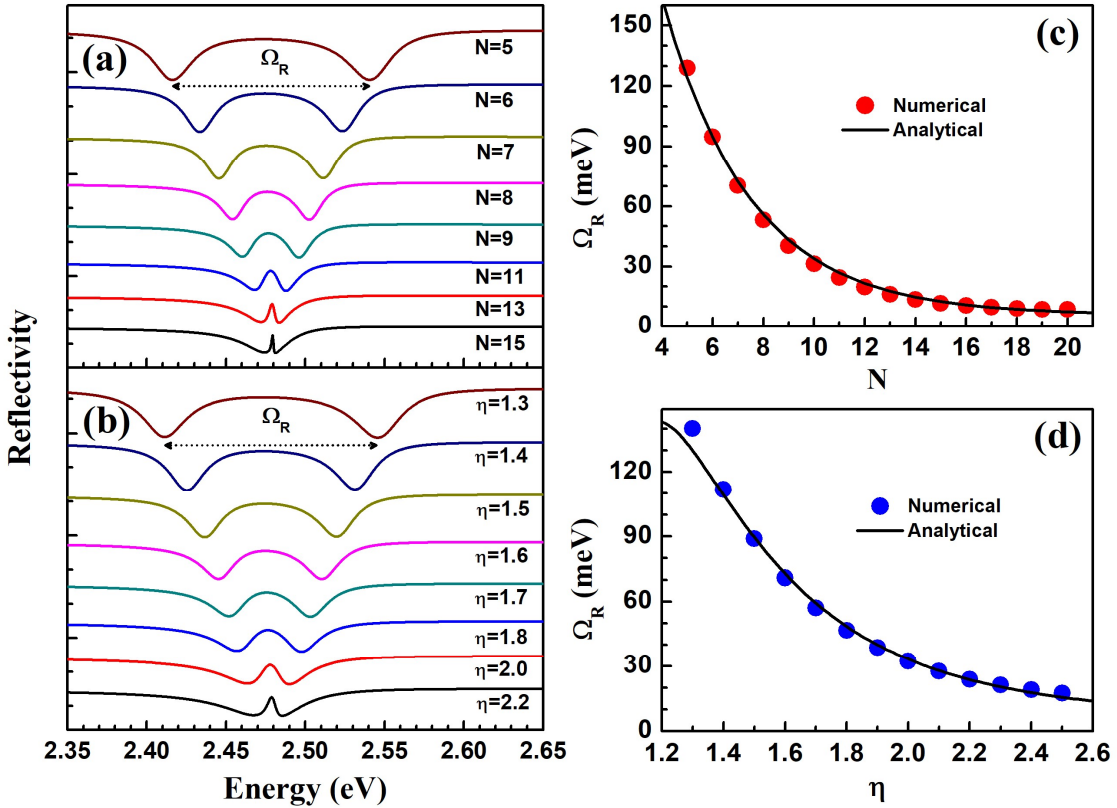

**Fig. 5.** Reflection spectra of the hybrid photonic structure **(a)** with varying number of period $N$ for a fixed value of $\eta$=1.333, and **(b)** with varying refractive index contrast $\eta$ for a fixed value of $N$=5. The energy separation between the two dips marked with a double-arrow line is the Rabi-like splitting energy ($\Omega_R$). Variation of $\Omega_R$ as a function of **(c)** $N$, and **(d)** $\eta$. The solid black line curves (———) in both the plot c and d are calculated using equation (31) with varying $N$ and $\eta$, respectively.

$$\Omega_{cT} = \frac{1}{\pi}\left(\frac{1}{\eta}\right)^N \left[\frac{1-(1/\eta)}{\sqrt{1-(1/2\eta)}}\right]\omega_0 \tag{30}$$

The Rabi-like splitting energy ($\Omega_R$) is double of $\Omega_{cT}$ and is given by

$$\Omega_R = 2\Omega_{cT} = \frac{2}{\pi}\left(\frac{1}{\eta}\right)^N \left[\frac{1-(1/\eta)}{\sqrt{1-(1/2\eta)}}\right]\omega_0 \tag{31}$$

where $\eta=n_H/n_L$ is the refractive index contrast, and $N$ is the number of periods in the 1DPC. Equation (31) is independent of properties of metallic layer. The analytically obtained values of $\Omega_R$ using equation (31) for varying $N$ and $\eta$ are exactly matching with that of numerically obtained values as shown in Fig. 5(c) and (d), respectively. The Rabi energy can be reach up to 125 meV for $N$=5, which exhibits strong coupling. For lower $N$ value, the resonant electric field corresponding to the TP mode easily penetrates through the 1DPC, and efficiently interacts with the cavity mode resulting in a large value of Rabi-like splitting energy. There is an exchange of energy between the cavity layer and metallic silver interface (close to S layer) because of strong coupling of cavity and TP modes, as a result an enhanced light-matter interaction occurs in either of the C and S layers. Strong interaction between cavity mode and TP mode can also be made by choosing lower $\eta$ value for a fixed value of $N$, which will subsequently give large value of $\Omega_R$ as shown in Fig. 5(d). The value of $\Omega_R$ decreases with increasing $\eta$ for the hybrid photonic structure. The effect of $\eta$ on the interaction of the cavity and TP modes in the coupled state is exactly identical to that of the effect of $N$. The role of $N$ and $\eta$ in a hybrid structure are interrelated and their combination can be optimized as per the desired interaction of the modes in the strong coupling region.

### 3.3. Polarization splitting

The variation of hybrid TP-cavity modes with varying angle of incidence for both TE and TM polarized light has been numerically estimated and plotted in Fig. 6(a) along with their bare modes. The hybrid modes ($\omega_U$ and $\omega_L$) are seen to stay away from each other over the entire angle of incidence unlike to their bare modes, which ensures the strong coupling between the cavity and TP modes. It is found that both the hybrid modes $\omega_U$ and $\omega_L$ exhibit identical parabolic dispersion curve irrespective of polarization of light. The resonant hybrid modes undergo blueshift to the higher energy with increasing angle of incidence. This is because the resonant energy of the modes must increase to maintain a fixed phase shift ($\delta$) with increasing angle ($\theta$) following the resonance condition $\delta = 2\pi nd\cos\theta/\lambda$, where $d$ is the virtual cavity thickness. The value of resonant reflectivity dip and its bandwidth of both the modes remain constant over a wider angle range up to nearly 20°. The value of reflectivity dip of the TM-polarized light for the hybrid modes remain unchanged over a wide angular range, which makes it useful for developing dual-narrow-band with tunable reflection in this region. The polarization splitting (TM-TE) for both the modes increase quadratically with increasing angle of incidence as shown in Fig. 6(b). The value of polarization splitting is higher for low energy mode $\omega_L$ as compared to that of $\omega_U$ and it becomes larger towards high angles of incidence. The estimated splitting value reaches up to 114 meV for $\omega_L$ and 9 meV for $\omega_U$, respectively at angle of $\theta=70°$. It is seen that the polarization splitting can be tuned by varying either thickness of metallic layer or spacing layer or both [42]. The polarization dependent coupling of TP and cavity modes provides better understandings about the mode structure. It shows that sharp hybrid resonant modes of orthogonal polarization can be tuned as per the requirement, which could be useful for optoelectronic device applications.

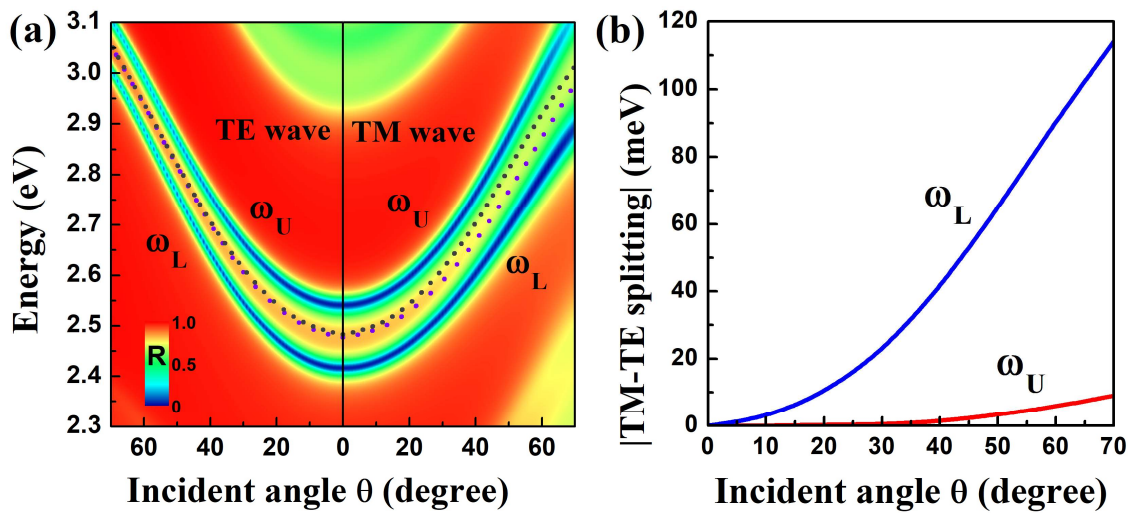

**Fig. 6.** (a) Reflectivity contour as a function of energy and incident angle for TE and TM-polarized light, respectively. The black and purple dotted lines on the contour plot represent the bare cavity and the bare TP mode, respectively. (b) Splitting between TE and TM-polarizations as a function of incident angle for the hybrid TP-cavity modes.

### 3.4. Refractive index sensing

The hybrid modes in the plasmonic-photonic crystal heterostructure can be utilized for refractive index sensing applications considering the fact that at least one of the modes must be sensitive to the change in the cavity medium. The objective is to sense refractive index values from 1.33 to 1.49 using the hybrid modes as this range covers the refractive indices of most of the biochemical samples, and therefore can be used for biochemical sensing. In this context, a sensing configuration has been designed in which two photonic structures PS1 and PS2 are kept at submicron distance apart forming a cavity as shown in Fig. 7 (a). The structure of PS1 is substrate/Ag/S/1DPC, while that of PS2 is 1DPC/substrate. The cavity (sub-micron thick) is filled with analytes to be sensed. Here, it is noteworthy to mention that the spacer layer can also be used as a s sensing medium *i.e.* the fluid flow can be placed between the metal layer and the 1DPC. It is seen that the sensitivity of the sensor is almost same whether the cavity or the spacer layer acts as a fluid flow channel. An experimental set-up has been proposed whose schematic is shown in Fig. 7 (b). In this set-up, UV-VIS-NIR light from the light source is sent through multiple illumination fibers to the hybrid structure (PS1/cavity/PS2) and the reflected light is collected by another optical fiber in the center of the reflection probe which is coupled to a spectrometer. In recent years, fabrication of hybrid plasmonic-nanofluidic sensing devices has been proposed and demonstrated [43]. Nanogaps and nanopores are also explored for nanofluidic structures for sensing application [44]. Shih *et al*. [45] have recently shown mid-infrared plasmonic liquid sensing in nanometric space driven by capillary space without any external liquid driving power source. The fluid flow channels passing through the cavity [46-48] can be designed for nanometric region of liquid analyte sensing. More feasible methods for making nanometric fluid flow channel could be possible with advances in nanofabrication technology in the near future.

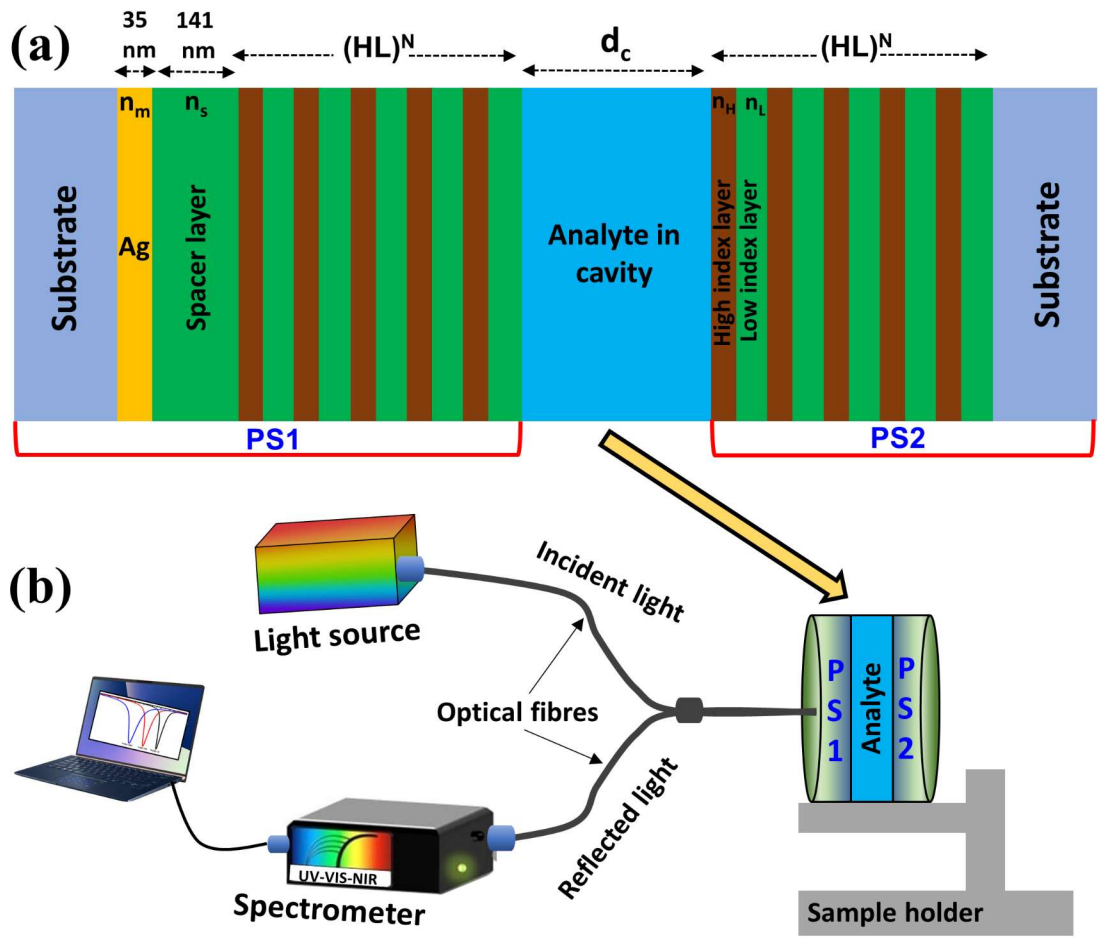

**Fig. 7: (a)** Schematic of hybrid photonic structure as a refractive index sensor. It consists of an analyte medium in the cavity between two photonic structures PS1 and PS2. **(b)** Proposed experimental set-up for sensing refractive index of the analyte medium.

Calculations have been carried out with this configuration to assess how sensitive the hybrid modes are with varying analyte medium. It is found that the designed hybrid plasmonic-photonic structure can give a good sensitivity and performance of the sensors, which is mentioned step by step as follows. At first, the behavior of resonant modes with varying cavity (analyte) thickness has been investigated for two hybrid structures S1 and S2 having different values of $\eta$ and $N$ in the 1DPCs. The structure S1 is designed as Ag/SiO$_2$/(HfO$_2$/SiO$_2$)$^{10}$/Analyte/(HfO$_2$/SiO$_2$)$^{10}$/Glass with $\eta=1.333$ ($n_{HfO2}/n_{SiO2}=1.96/1.47$) and $N=10$. The structure S2 is designed as Ag/SiO$_2$/(TiO$_2$/SiO$_2$)$^{6}$/Analyte/(TiO$_2$/SiO$_2$)$^{6}$/Glass

with $\eta=1.605$ ($n_{TiO_2}/n_{SiO_2}=2.36/1.47$) and $N=6$. The value of $N$ in both structures is optimized to obtain sharp resonant mode which is sensitive to the change in the analyte medium. The reflectivity contour of S1 structure as a function of wavelength and cavity thickness for two extreme analyte refractive index values $n_a=1.33$ and 1.49 are plotted in Fig. 8. For S2 structure, the reflectivity contours are provided in the supplementary material (see Fig. S5). The figure shows the strong coupling between the TP mode and cavity mode as evident from the anticrossing or splitting of the modes at different values of cavity thickness. For cavity mode to resonate near Bragg frequency $\omega_0$ at these thickness values, the light should have maximum transmission in the cavity structure which leads to the condition: $r_{CP}e^{2i\varphi_c}r_{CP}=1$, where $r_{CP}$ is the reflection co-efficient amplitude at cavity/1DPC interface, and $\varphi_c = n_c d_c \omega/c$ is the phase shift of light having frequency $\omega$ passing through the cavity. At $\omega \approx \omega_0$, that condition will lead to the equation $d_c = (m-1)\pi c/(2n_c\omega_0)$. For cavity medium having refractive index $n_c=1.33$ (1.49), the values of $d_c \approx$ 94 nm (84 nm), 282 nm (252 nm), 470 nm (420 nm), 658 nm (588 nm), 846 nm (756 nm),………, respectively for $m=2, 4, 6, 8, 10,$………, which exactly matches to the anti-crossing $d_c$ values observed in Fig. 8. It can be seen that the resonant mode $\lambda_a$ significantly varies with analyte thickness, while that of $\lambda_R$ almost remains unchanged except at and around the strong coupling region as observed in the Fig. 8. From sensing point of view, one should avoid the cavity thickness values for which the strong coupling occurs. Therefore, the five values of cavity thickness ($d_c$=170 nm, 340 nm, 510 nm, 705 nm, and 892 nm) for which one mode remains fixed while other modes change largely has been chosen for refractive index sensing. There exist multiple analyte modes for a given value of $d_c$ as marked by arrows: one is above and other is below $\lambda_R$. One can use either of them or both for refractive index sensing.

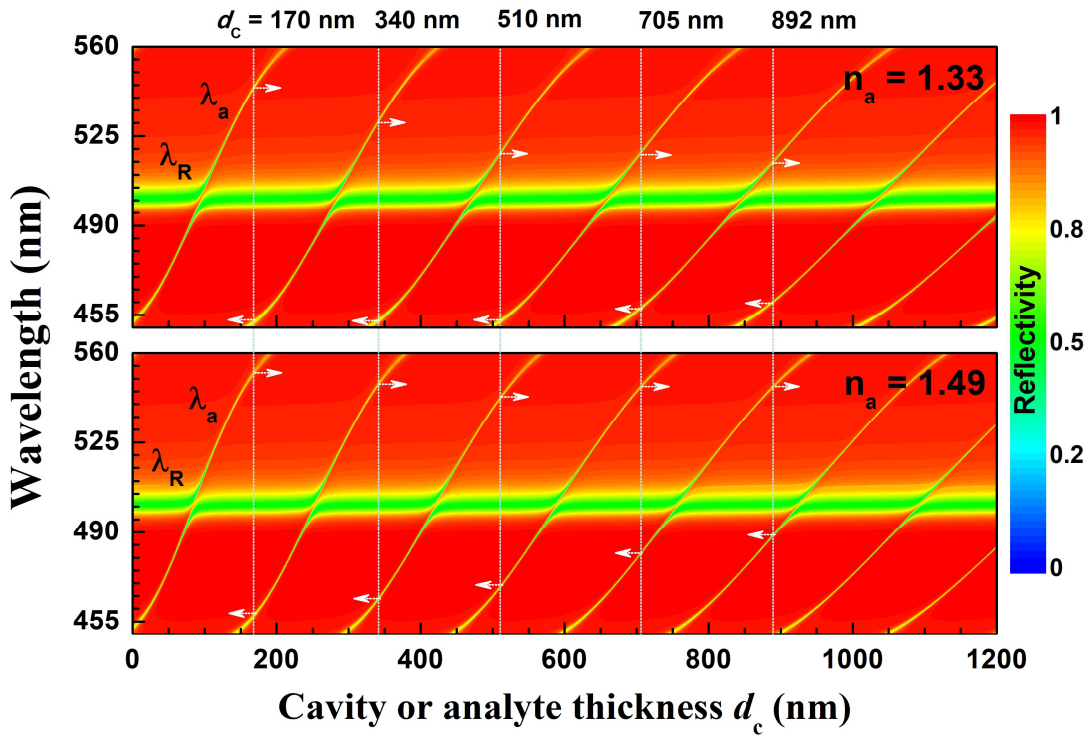

**Fig. 8.** Reflectivity contour as a function of wavelength and cavity or analyte thickness $d_c$ for two different values of refractive index of analyte: $n_a$ = 1.33 and 1.49, respectively in a hybrid structure of MS(HL)$^N$C(HL)$^N$ with (HL)$^N$ = (HfO$_2$/SiO$_2$)$^{10}$. The cavity layer C has been filled with analyte. The modes splitting has been found at the strong coupling region for different values of cavity thickness. For cavity thickness values other than strong coupling region, the reference mode $\lambda_R$ remains unchanged, while the analyte mode $\lambda_a$ shifts. Such cavity thickness values can be selected for self-referenced sensing applications. Here, 5-values of cavity thickness $d_c$=170 nm, 340 nm, 510 nm, 705 nm, and 892 nm, which lie in the weak coupling region, are selected as examples. The arrow indicated towards right side in the plot represents $\lambda_a$ above $\lambda_R$, while the arrow directed towards left side represents $\lambda_a$ below $\lambda_R$. One can clearly see the shifting of both lower and upper analyte mode with varying refractive index of analyte from 1.33 to 1.49 for the selected values of cavity thickness, which is explored for refractive index sensing application.

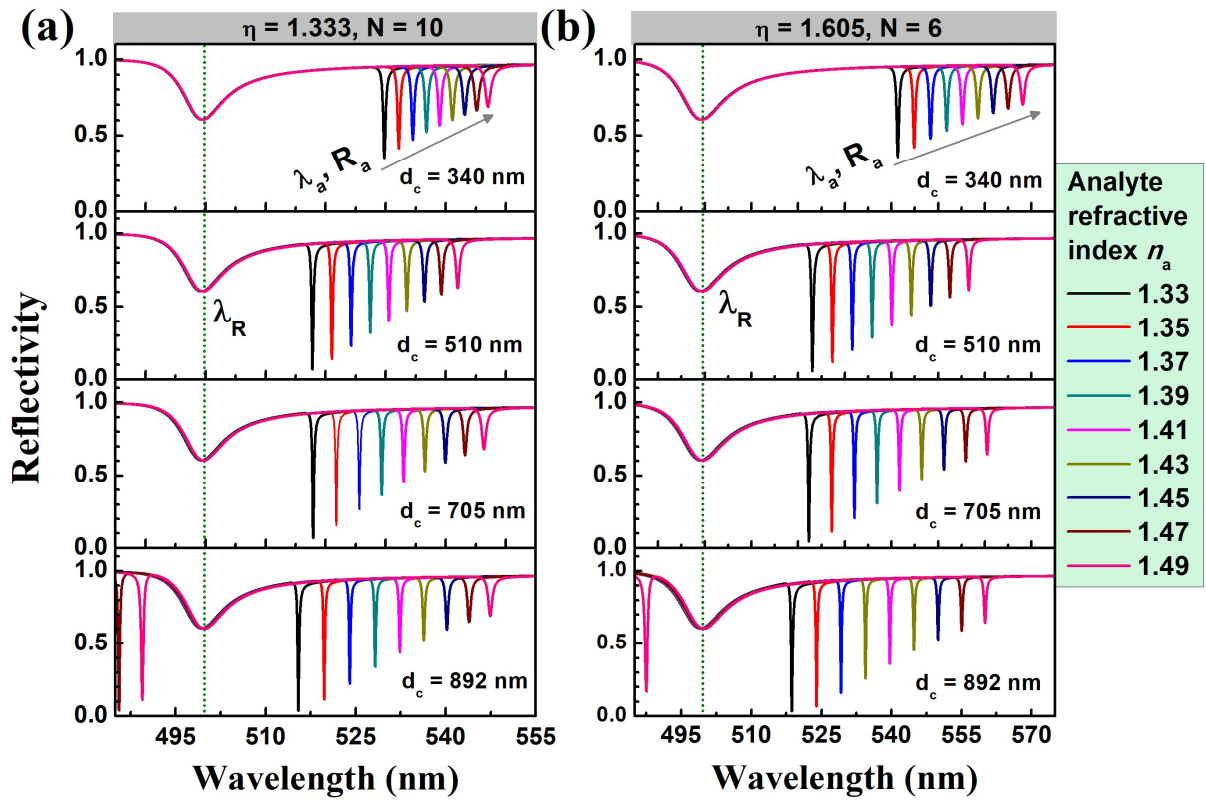

**Fig. 9.** Reflectivity spectra with varying refractive index values of analyte filled in the cavity in the hybrid structures MS(HL)$^N$C(HL)$^N$ with (HL)$^N$ as **(a)** (HfO$_2$/SiO$_2$)$^{10}$ 1DPC with refractive index contrast $\eta$=1.333 and number of periods $N$=10 [structure S1], and **(b)** (TiO$_2$/SiO$_2$)$^6$ 1DPC with $\eta$=1.605 and $N$=6 [structure S2], for different values of cavity or analyte thickness: $d_c$= 340 nm, 510 nm, 705 nm, and 892 nm.

In the present study, we have used analyte resonant mode $\lambda_a$ above $\lambda_R$ for refractive index sensing application. Subsequently, the reflectivity spectra for both structures S1 and S2 with varying refractive index of analytes in the range 1.33-1.49 have been calculated for the above mentioned five different cavity thickness values and plotted in Fig. 9(a) and (b), respectively. The wider wavelength range reflectivity spectra that exhibit multiple analyte modes (one above $\lambda_R$ and other below $\lambda_R$) with varying refractive index of analyte and cavity thickness are provided in the supplementary material [see Fig. S6]. It can be clearly seen that the reflectivity dip due to the mode $\lambda_R$ remains unchanged with varying analyte refractive index. On the contrary, the reflectivity dip ($R_a$) at the resonant analyte mode $\lambda_a$ undergoes significant shift as a function of refractive index of the analytes filled in the cavity. This makes it useful for designing "self-referenced sensor", where $\lambda_R$ becomes the reference mode and remains fixed, while $\lambda_a$ shifts considerably with varying optical property of the cavity medium. It is worthy to note that the shift of $\lambda_a$ with respect $\lambda_R$ becomes larger with increasing cavity thickness $d_c$ in both structures S1 and S2. It is found that further increase of cavity thickness generates additional resonant modes within the PBG, consequently the design of the sensor becomes complex. Moreover, this could lead to shift of $\lambda_a$ near the edge of the PBG that would badly affect the signal to noise ratio (SNR) of the sensor. Therefore, only five optimized cavity thickness values are chosen for the present proposed sensors.

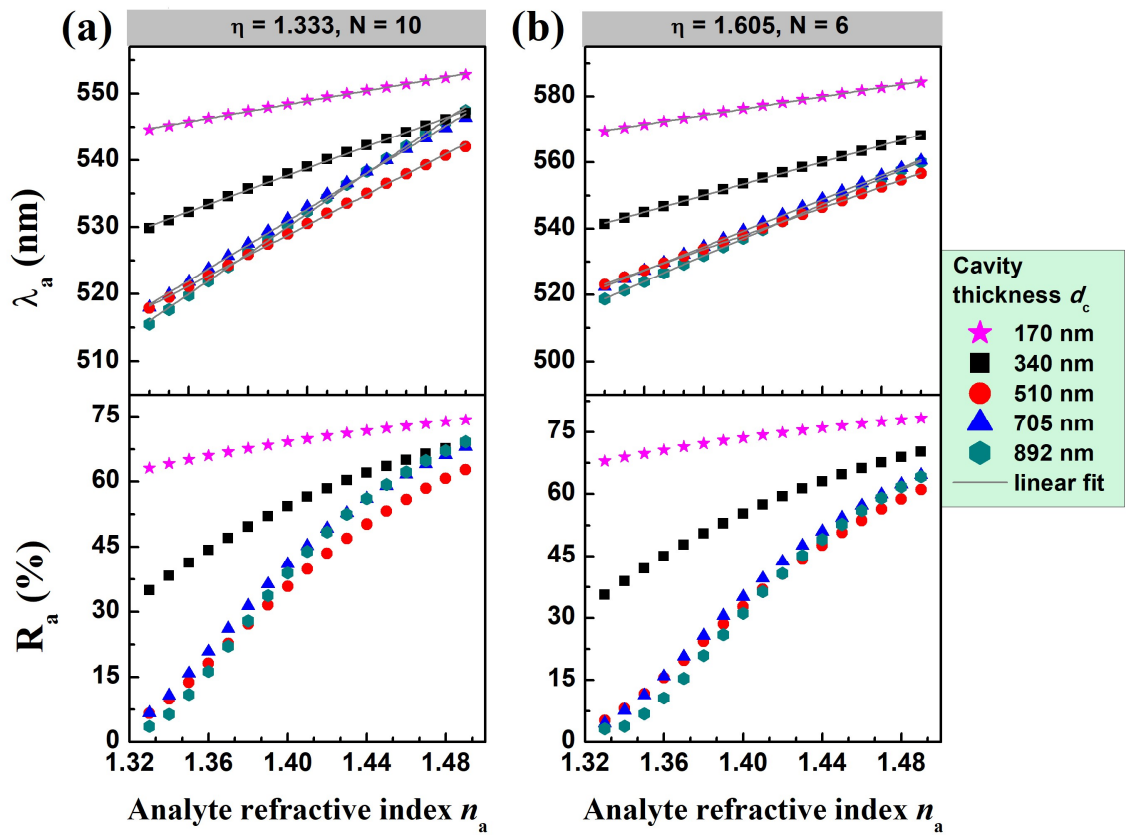

**Fig. 10.** Position of analyte mode wavelength ($\lambda_a$) and reflectivity minimum ($R_a$) at $\lambda_a$ as a function of analyte refractive index for the hybrid structures (a) S1 and (b) S2 for different cavity thicknesses.

The values of $\lambda_a$ and $R_a$ with varying analyte medium for different cavity thickness are plotted in Fig. 10 (a) and (b) for the structures S1 and S2, respectively. The value of $\lambda_a$ shifts to longer wavelength and the value of reflectivity dip $R_a$ increases with increasing refractive index of the analyte which suggests that either of change in $\lambda_a$ and $R_a$ or both can be used for the refractive index sensing application. The change in both $\lambda_a$ and $R_a$ is maximum for cavity thickness of 892 nm in both the structures. Such significant changes could be understood by analyzing the electric field distribution of $\lambda_a$. For better visualization and understanding of electric field intensity in the hybrid structure, the cavity or analyte layer center is considered as z=0 for the electric field calculations. The analyte mode electric field distribution has been plotted for different values of analyte index $n_a$=1.35, 1.40, and 1.45 in Fig. 11(a) for a fixed value of $d_c$=340 nm and $\eta$=1.333. This shows that the electric field within the analyte layer around z=0 gets enhanced by nearly four times with decreasing the analyte index from 1.45 to 1.35, which indicates that the cavity/analyte mode would exhibit better quality factor for low index analyte as compared to that of high index analyte. The analyte mode field distribution with varying thickness of cavity or analyte layer has been plotted in Fig. 11(b) for a fixed value of $n_a$=1.35 and $\eta$=1.333. It shows that the field gets much stronger and more localized with increasing cavity thickness from 170 nm to 510 nm. As a result, the bandwidth of the analyte mode will be narrower and the dip reflectivity will be deeper with increasing cavity thickness. This in turn improves the sensitivity and quality factor of the analyte mode. Fig. 11(c) shows the mode field distribution for two different values of refractive index contrast $\eta$=1.333 and 1.605 in the 1DPC. The mode field is much stronger and confined in higher index contrast photonic structure. The electric field intensity becomes ten times by increasing index contrast from 1.333 to 1.605, which confirms the increasing sensitivity of the analyte modes in case of higher index contrast structure. The change in electric field distribution in all the considered cases is significant, which is essentially the prime reason behind such a large shift of the resonant analyte mode $\lambda_a$ as observed in Fig. 9. In all cases, the electric field is seen amplified in both 1DPC structures also, besides the amplification in the cavity. The presence of metallic layer and 1DPC makes the whole structure similar to a Fabry-Perot cavity resulting in resonances both in the cavity and in the 1DPCs. Such field amplification leads to high Q value for the coupled modes.

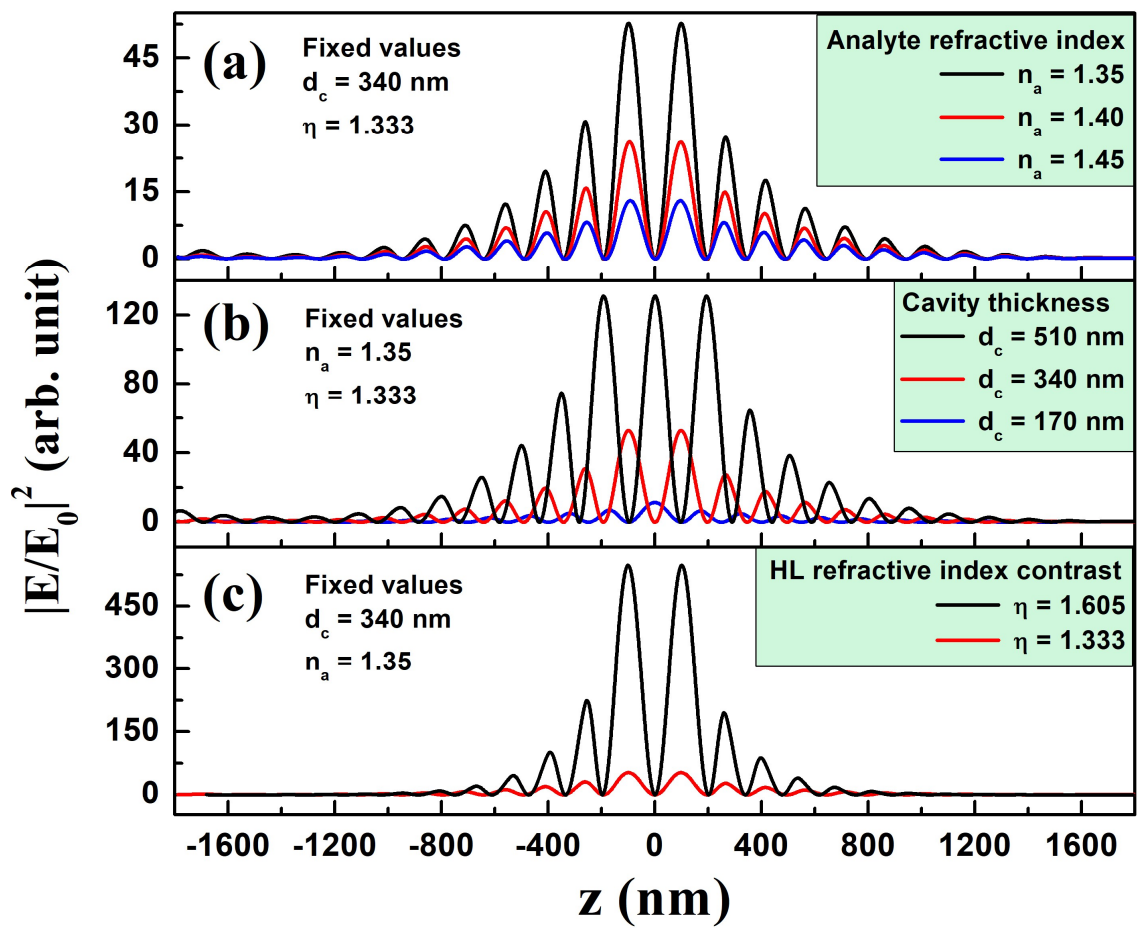

**Fig. 11.** Electric field distribution of the analyte modes in the hybrid structure for **(a)** different refractive index values of the analyte $n_a$ with a fixed value of $d_c$ and $\eta$, **(b)** for different values of the cavity thickness $d_c$ with a fixed value of $n_a$ and $\eta$, and **(c)** for different values of refractive index of the 1DPC $\eta$ with a fixed value of $d_c$ and $n_a$.

Now, the performance of the hybrid modes-based sensor for both the structure S1 and S2 has been evaluated by analyzing the shift of $\lambda_a$. The sensitivity $S_n$ of the sensor is defined as $S_n = \Delta\lambda_a / \Delta n_a$, where $\Delta\lambda_a$ is the shift of the analyte mode and $\Delta n_a$ is the change in refractive index of the analyte (cavity medium). Sensitivity has been obtained from the slope of the plots $\lambda_a$ vs. $n_a$ in Fig. 10. The derived sensitivity of both structures S1 and S2 as a function of cavity thickness are shown in Fig. 12 (a). The sensitivity is found increasing from 51 nm/RIU to 201 nm/RIU and 93 nm/RIU to 259 nm/RIU with increasing cavity thickness from 170 nm to 892 nm for the structures S1 and S2, respectively. The sensitivity obtained for both the structures is better than that of the previously reported values [23, 26] for the TP based sensors. The structure S2 has better sensitivity than S1. This clearly indicates that hybrid plasmonic-1D photonic crystal structure made of dielectric materials with wider refractive index contrast gives better sensitivity. Moreover, wider the index contrast $\eta$, lesser is the number of period $N$ required to achieve the desired PBG. The maximum sensitivity obtained for the structure S2 is 259 nm/RIU, which is much better than various photonic crystal-fiber based interferometric sensors [49], and recently reported TP mode based mesoporous multilayer sensor [50]. The sensitivity of SPR based sensor is better compared to the proposed hybrid structure-based sensor primarily due to polarization dependent strong dispersive nature of SPR mode. However, the sensing configuration of hybrid mode-based sensor is much easier to realise than that of SPR sensors.

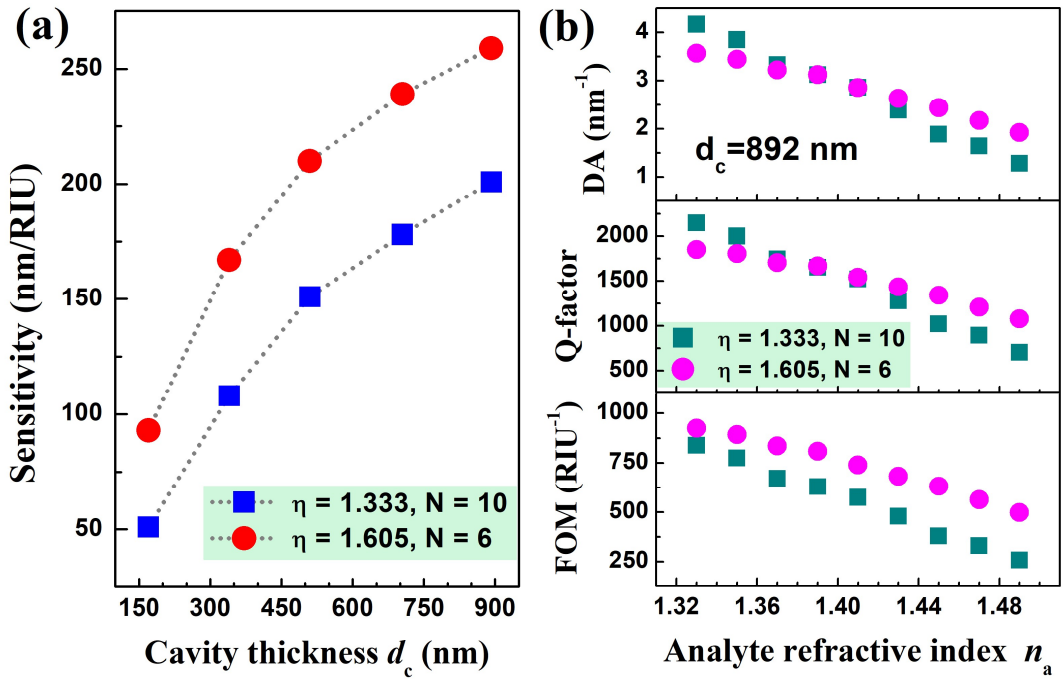

**Fig. 12.** (a) Sensitivity of the hybrid photonic structures S1 and S2 as a function of cavity thickness. (b) DA, Q-factor, and FOM for both the structures as a function of refractive index of the analyte for a cavity thickness of 892 nm.

Other than sensitivity, detection accuracy (*DA*), quality factor (*Q*-factor), and figure of merit (*FOM*) are also vital parameters that decide the performance of the sensor and are expressed as follows: $DA = 1/\delta_a$, $Q-factor = \lambda_a/\delta_a$, and $FOM = S_n/\delta_a$, where $\delta_a$ is the fullwidth at half minima of the reflection dip corresponding to the analyte mode $\lambda_a$. These expressions indicate that sharper resonant modes improve the overall performance of the sensors. The value of *DA*, *Q*-factor, and *FOM* has been found decreasing with increasing refractive index of the analyte. The variation of such parameters for sensor structures S1 and S2 with maximum sensitivity (for cavity thickness of 892 nm) are plotted in Fig. 12(b) as a representative plot. The value of *DA* decreases from 4.16 nm$^{-1}$ to 1.28 nm$^{-1}$ for S1, and 3.57 nm$^{-1}$ to 1.92 nm$^{-1}$ for S2 with increasing $n_a$ from 1.33 to 1.49. The obtained *DA* is much better than that obtained in case of SPR sensors. The typical *DA* for SPR sensors is around 0.02 nm$^{-1}$ in the visible region which is due to the involvement of additional dispersion compensating components (prisms and gratings) for exciting the SPR mode. The Q-factor decreases from 2148 to 701 for S1, and 1852 to 1077 for S2 with increasing $n_a$ from 1.33 to 1.49. The FOM decreases from 837 RIU-1 to 258 RIU$^{-1}$ for S1, and 925 RIU-1 to 498 RIU$^{-1}$ for S2 with increasing analyte index in the given range. It is observed that the DA and Q-factor of the sensor made of structure S1 is better than that of S2 for the refractive index of the analyte in the range 1.33 to 1.37. However, it depicts reverse behaviour for $n_a$ >1.37. The DA and Q-factor of structure S2 can be better than that of S1 in the whole range of refractive index of the analyte if larger value of *N* is chosen. Overall, structure S2 exhibits better sensing performance than that of S1. It is worth to mention here that the value of *DA*, *FOM*, and *Q*-factor of the structure S2 could be further improved by increasing the number of periods N as it will drastically narrow the FWHM of the reflection dip. This confirms that the performance of the hybrid plasmonic-photonic structure-based sensors can be further improved by optimizing number of periods *N* and/or thickness of metal and spacer layers, and choosing high refractive index contrast η of dielectric materials for 1DPC. This proposed hybrid photonic structure will be much easier to configure for a sensor as the hybrid modes are polarization insensitive and the light can be easily coupled to the structure in free space to excite the modes for refractive index sensing application.

## 4. Conclusion

Rabi-like splitting and self-referenced refractive index sensing in Ag/S/1DPC/cavity/1DPC structures have been realized using hybrid TP-cavity modes. Anticrossing of the TP mode and cavity mode has been observed at specific values of spacer layer thickness, which is explained analytically considering the Ag/S/1DPC as a Fabry-Perot cavity like structure. The coupling between the two modes in the hybrid structure has been explained using a coupled oscillator model. It is found that Rabi-like splitting energy is tunable by either varying *N* for a fixed *η* or varying *η* for a fixed *N*. The polarization splitting of one of the hybrid modes ($\omega_L$) is much higher as compared to that of the other mode ($\omega_U$). The modes coupling changes by varying thickness and refractive index of the analyte medium in the cavity in which the value of resonant reflectivity and the corresponding wavelength of one of the modes remains unchanged while that of the other mode undergoes significant change, therefore the structure acts as a self-referenced refractive index sensor for a fixed cavity thickness. The response of the two proposed hybrid structure sensors is linear for the analytes in the refractive index range of 1.33 to 1.49. The structures made of refractive index contrast $\eta=1.605$ and $N=6$ shows better sensing performance as compared to that made of $\eta = 1.333$ and $N=10$. The sensitivity of the sensor ($\eta=1.605$ and $N=6$) is comparable to that of fiber interferometer sensors and SPR sensors. Moreover, the thickness of cavity (sensing) medium in the proposed structure has been made close to 900 nm which is much larger than that of the reported values so far for the TP based sensors. The larger thickness of sensing medium not only improves its sensitivity but also makes the hybrid structure more practically feasible. The proposed structures can be used for chemical or biochemical sensing by integrating nano/microfluidic channel in the hybrid plasmonic-photonic crystals.

# References

[1] M. Pu, X. Ma, Y. Guo, X. Li, X. Luo 2018 Theory of microscopic meta-surface waves based on catenary optical fields and dispersion *Opt. Express* **26** 19555-19562

[2] Z.A. Zaky, A.M. Ahmed, A.S. Shalaby, A.H. Aly 2020 Refractive index gas sensor based on the Tamm state in a one-dimensional photonic crystal: theoretical optimisation *Sci. Rep.* **10** 1-9

[3] Z. Wang, J.K. Clark, Y.-L. Ho, S. Volz, H. Daiguji, J.-J. Delaunay 2020 Ultranarrow and wavelength-tunable thermal emission in a hybrid metal–optical tamm state structure *ACS Photonics* **7** 1569-1576

[4] M. Kaliteevski, I. Iorsh, S. Brand, R. Abram, J. Chamberlain, A. Kavokin, I. Shelykh 2007 Tamm plasmon-polaritons: Possible electromagnetic states at the interface of a metal and a dielectric Bragg mirror *Phys. Rev. B* **76** 165415

[5] M. Sasin, R. Seisyan, M. Kalitteevski, S. Brand, R. Abram, J. Chamberlain, A.Y. Egorov, A. Vasil'Ev, V. Mikhrin, A. Kavokin 2008 Tamm plasmon polaritons: Slow and spatially compact light *Appl. Phys. Lett.* **92** 251112

[6] P.S. Pankin, S.Y. Vetrov, I.V. Timofeev 2017 Tunable hybrid Tamm-microcavity states *J. Opt. Soc. Am. B* **34** 2633-2639

[7] M.V. Pyatnov, S.Y. Vetrov, I.V. Timofeev 2017 Localized optical modes in a defect-containing liquid-crystal structure adjacent to the metal *J. Opt. Soc. Am. B* **34** 2011-2017

[8] Z. Wang, J.K. Clark, Y.-L. Ho, B. Vilquin, H. Daiguji, J.-J. Delaunay 2018 Narrowband thermal emission from Tamm plasmons of a modified distributed Bragg reflector *Appl. Phys. Lett.* **113** 161104

[9] A.M. Ahmed, A. Mehaney 2019 Ultra-high sensitive 1D porous silicon photonic crystal sensor based on the coupling of Tamm/Fano resonances in the mid-infrared region *Sci. Rep.* **9** 1-9

[10] P. Das, S. Mukherjee, M. Wan, S.K. Ray 2018 Optical Tamm state aided room-temperature amplified spontaneous emission from carbon quantum dots embedded one-dimensional photonic crystals *J. Phys. D: Appl. Phys.* **52** 035102

[11] N. Lundt, S. Klembt, E. Cherotchenko, S. Betzold, O. Iff, A.V. Nalitov, M. Klaas, C.P. Dietrich, A.V. Kavokin, S. Höfling 2016 Room-temperature Tamm-plasmon exciton-polaritons with a WSe 2 monolayer *Nat. Commun.* **7** 1-6

[12] J. Hu, E. Yao, W. Xie, W. Liu, D. Li, Y. Lu, Q. Zhan 2019 Strong longitudinal coupling of Tamm plasmon polaritons in graphene/DBR/Ag hybrid structure *Opt. Express* **27** 18642-18652

[13] Z. Wang, J.K. Clark, Y.-L. Ho, B. Vilquin, H. Daiguji, J.-J. Delaunay 2018 Narrowband thermal emission realized through the coupling of cavity and Tamm plasmon resonances *ACS Photonics* **5** 2446-2452

[14] H. Lu, Y. Li, H. Jiao, Z. Li, D. Mao, J. Zhao 2019 Induced reflection in Tamm plasmon systems *Optics Express* **27** 5383-5392

[15] R.V. Nair 2019 The interaction between optical Tamm state and microcavity mode in a planar hybrid plasmonic-photonic structure *Photonics Nanostructures - Fundam. Appl.* **36** 100702

[16] F.H. Alast, G. Li, K. Cheah 2017 Rabi-like splitting from large area plasmonic microcavity *AIP Advances* **7** 085201

[17] G. Khitrova, H. Gibbs, M. Kira, S.W. Koch, A. Scherer 2006 Vacuum Rabi splitting in semiconductors *Nature Physics* **2** 81-90

[18] L. Zhang, Y. Zhang, Y. Zhao, J. Zhai, L. Li 2010 Rabi splitting induced by a metamaterial plasmon cavity *Opt. Express* **18** 25052-25060

[19] S. Chen, G. Li, D. Lei, K.W. Cheah 2013 Efficient energy exchange between plasmon and cavity modes via Rabi-analogue splitting in a hybrid plasmonic nanocavity *Nanoscale* **5** 9129-9133

[20] P. Das, S. Mukherjee, S. Jana, S.K. Ray, B.S. Bhaktha 2020 Resonant and non-resonant coupling of one-dimensional microcavity mode and optical Tamm state *J. Opt.* **22** 065002

[21] R. Brückner, M. Sudzius, S. Hintschich, H. Fröb, V. Lyssenko, K. Leo 2011 Hybrid optical Tamm states in a planar dielectric microcavity *Phys. Rev. B* **83** 033405

[22] M. Maragkou, C.E. Richards, T. Ostatnický, A.J. Grundy, J. Zajac, M. Hugues, W. Langbein, P.G. Lagoudakis 2011 Optical analogue of the spin Hall effect in a photonic cavity *Opt. Lett.* **36** 1095-1097

[23] S. Kumar, M.K. Shukla, P.S. Maji, R. Das 2017 Self-referenced refractive index sensing with hybrid-Tamm-plasmon-polariton modes in sub-wavelength analyte layers *J. Phys. D: Appl. Phys.* **50** 375106

[24] C. Symonds, A. Lemaître, E. Homeyer, J. Plenet, J. Bellessa 2009 Emission of Tamm plasmon/exciton polaritons *Appl. Phys. Lett.* **95** 151114

[25] E. Gonzalez-Valencia, I. Del Villar, P. Torres 2020 Bloch waves at the surface of a single-layer coating D-shaped photonic crystal fiber *Opt. Lett.* **45** 2547-2550

[26] P.S. Maji, M.K. Shukla, R. Das 2018 Blood component detection based on miniaturized self-referenced hybrid Tamm-plasmon-polariton sensor *Sens. Actuators B Chem.* **255** 729-734

[27] E. Gonzalez-Valencia, R.A. Herrera, P. Torres 2019 Bloch surface wave resonance in photonic crystal fibers: towards ultra-wide range refractive index sensors *Opt. Express* **27** 8236-8245

[28] A.L. Lereu, M. Zerrad, A. Passian, C. Amra 2017 Surface plasmons and Bloch surface waves: Towards optimized ultra-sensitive optical sensors *Appl. Phys. Lett.* **111** 011107

[29] M. Shaban, A.M. Ahmed, E. Abdel-Rahman, H. Hamdy 2017 Tunability and sensing properties of plasmonic/1D photonic crystal *Sci. Rep.* **7** 1-10

[30] L.A. Pettersson, L.S. Roman, O. Inganäs 1999 Modeling photocurrent action spectra of photovoltaic devices based on organic thin films *J. Appl. Phys.* **86** 487-496

[31] C.C. Katsidis, D.I. Siapkas 2002 General transfer-matrix method for optical multilayer systems with coherent, partially coherent, and incoherent interference *Appl. Opt.* **41** 3978-3987

[32] E. Centurioni 2005 Generalized matrix method for calculation of internal light energy flux in mixed coherent and incoherent multilayers *Appl. Opt.* **44** 7532-7539

[33] H.A. Macleod 2017 *Thin-film optical filters* DOI: CRC press)

[34] S. Larouche, L. Martinu 2008 OpenFilters: open-source software for the design, optimization, and synthesis of optical filters *Appl. Opt.* **47** C219-C230

[35] P. Peumans, A. Yakimov, S.R. Forrest 2003 Small molecular weight organic thin-film photodetectors and solar cells *J. Appl. Phys.* **93** 3693-3723

[36] J. Hu, W. Liu, W. Xie, W. Zhang, E. Yao, Y. Zhang, Q. Zhan 2019 Strong coupling of optical interface modes in a 1D topological photonic crystal heterostructure/Ag hybrid system *Opt. Lett.* **44** 5642-5645

[37] Y.-t. Fang, L.-x. Yang, W. Kong, N. Zhu 2013 Tunable coupled states of a pair of Tamm plasmon polaritons and a microcavity mode *J. Opt.* **15** 125703

[38] M. Born, E. Wolf 2013 *Principles of optics: electromagnetic theory of propagation, interference and diffraction of light* DOI: Elsevier)

[39] B.J. Lee, Z. Zhang 2006 Design and fabrication of planar multilayer structures with coherent thermal emission characteristics *J. Appl. Phys.* **100** 063529

[40] H.U. Yang, J. D'Archangel, M.L. Sundheimer, E. Tucker, G.D. Boreman, M.B. Raschke 2015 Optical dielectric function of silver *Phys. Rev. B* **91** 235137

[41] M. Kaliteevski, S. Brand, R. Abram, I. Iorsh, A. Kavokin, I. Shelykh 2009 Hybrid states of Tamm plasmons and exciton polaritons *Appl. Phys. Lett.* **95** 251108

[42] R. Brückner, M. Sudzius, S. Hintschich, H. Fröb, V. Lyssenko, M. Kaliteevski, I. Iorsh, R. Abram, A. Kavokin, K. Leo 2012 Parabolic polarization splitting of Tamm states in a metal-organic microcavity *Appl. Phys. Lett.* **100** 062101

[43] J. Xu, Z. Ren, B. Dong, X. Liu, C. Wang, Y. Tian, C. Lee 2020 Nanometer-scale heterogeneous interfacial sapphire wafer bonding for enabling plasmonic-enhanced nanofluidic mid-infrared spectroscopy *ACS Nano* **14** 12159-12172

[44] X. Miao, L. Yan, Y. Wu, P.Q. Liu 2021 High-sensitivity nanophotonic sensors with passive trapping of analyte molecules in hot spots *Light Sci. Appl.* **10** 1-11

[45] K. Shih, Z. Ren, C. Wang, C. Lee 2019 MIR plasmonic liquid sensing in nano-metric space driven by capillary force *J. Phys. D: Appl. Phys.* **52** 394001

[46] K.V. Sreekanth, Y. Alapan, M. ElKabbash, E. Ilker, M. Hinczewski, U.A. Gurkan, A. De Luca, G. Strangi 2016 Extreme sensitivity biosensing platform based on hyperbolic metamaterials *Nat. Mater.* **15** 621-627

[47] A. Salim, S. Lim 2018 Review of recent metamaterial microfluidic sensors *Sensors* **18** 232

[48] X. Liang, K.J. Morton, R.H. Austin, S.Y. Chou 2007 Single sub-20 nm wide, centimeter-long nanofluidic channel fabricated by novel nanoimprint mold fabrication and direct imprinting *Nano Lett.* **7** 3774-3780

[49] R. Jha, J. Villatoro, G. Badenes 2008 Ultrastable in reflection photonic crystal fiber modal interferometer for accurate refractive index sensing *Appl. Phys. Lett.* **93** 191106

[50] B. Auguié, M.C. Fuertes, P.C. Angelomé, N.L. Abdala, G.J. Soler Illia, A. Fainstein 2014 Tamm plasmon resonance in mesoporous multilayers: toward a sensing application *ACS Photonics* **1** 775-780

# Supplementary material

## Rabi-like splitting and refractive index sensing with hybrid Tamm plasmon-cavity modes

S. Jena[*], R. B. Tokas, S. Thakur, and D. V. Udupa

Atomic & Molecular Physics Division, Bhabha Atomic Research Centre, Mumbai 400 085, India

**Coupling of modes without spacing layer:**

Top plot of Fig. S1 shows the bare cavity mode in the structure $(HL)^5C(HL)^5$. The middle plot of Fig. S1 is the TP mode at lower energy region in the structure $M(HL)^{10}$, which is originated due to the electric field localized next to the metallic layer. Combining above two structures leads to $M(HL)^5C(HL)^5$ and it exhibits both the modes as shown in the bottom plot of Fig. S1. But one of the modes undergoes red-shift and it is the TP mode. Here, the modes are clearly distinguishable as they are poorly coupled. It can be clearly seen that the modes are weakly coupled in the absence of spacing layer. Therefore, spacing layer is essential for strong coupling of the modes as detailed in the main manuscript.

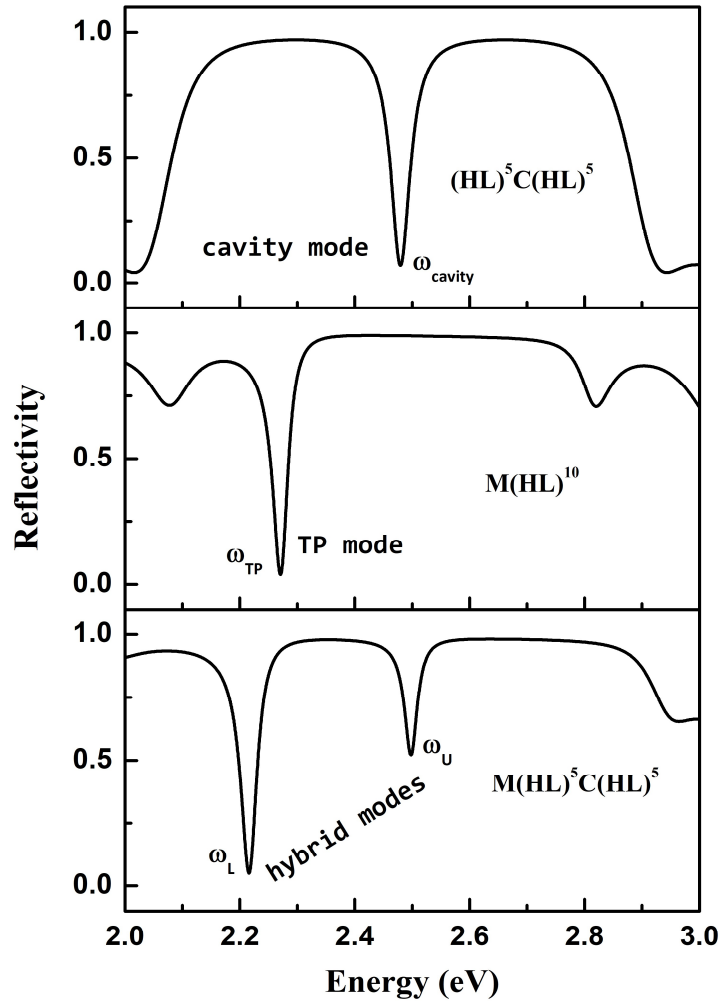

**Fig. S1.** Hybrid $M(HL)^5C(HL)^5$ structure without spacing S layer: Reflection spectra of the photonic structure without metallic M layer (bare cavity mode), without cavity C layer (bare TP mode), and with both M and C layers (hybrid TP-cavity modes), respectively.

## Optimization of plasmonic layer thickness:

Fig. S2 shows that TP mode exists in both the structures for a specific range of values of $d_m$. In case of $MS(HL)^{10}$, the $d_m$ values lies in the range 40-75 nm, while it is in the range 30-40 nm for the $MS(HL)^5 C(HL)^5$ hybrid photonic structure. Therefore, optimized value of $d_m$ should be chosen for the desired coupling of the TP mode with the cavity mode, which is 35 nm in the present study. Correlating Fig. S2(b) with Fig. S2(a), it can be inferred that the energy $\omega_U = 2.541$ eV corresponds to the cavity mode part whereas the energy $\omega_L = 2.416$ eV corresponds to TP mode part in the hybrid TP-cavity modes.

*High index thin layer $HfO_2$ [$n_H = 1.96$]
*Low index thin layer $SiO_2$ [$n_L = 1.47$]

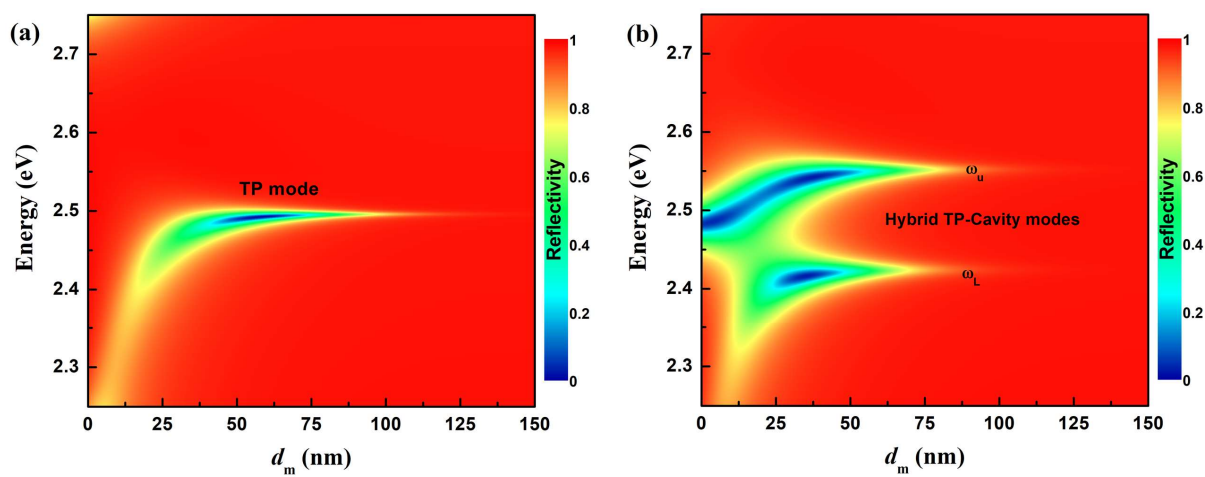

**Fig. S2.** Reflectivity spectra as a function of metallic M layer thickness ($d_m$) for **(a)** $MS(HL)^{10}$, and **(b)** $MS(HL)^5 C(HL)^5$ photonic structure, respectively.

For the structure made of $TiO_2$ ($n_H = 2.34$) and $SiO_2$ ($n_L = 1.47$), the numerically obtained range of $d_m$ values is 35 nm-55 nm for the ($TiO_2/SiO_2$) based hybrid plasmonic-photonic structure. Therefore, $d_m = 35$ is chosen for both the $HfO_2/SiO_2$ as well as $TiO_2/SiO_2$ based structures.

## Effect of spacing layer thickness on bare mode and hybrid modes:

Fig. S3 illustrates the effect of spacer layer thickness ($d_s$) on the bare modes and coupled hybrid modes. It is seen that the bare TP mode undergoes significant shift with increasing $d_s$ as shown in Fig. S3(a), while the bare cavity mode does not change with varying S layer thickness as shown in Fig. S3(b). In case of hybrid structure, the coupled modes deviate from their bare TP and cavity modes in the strong coupling region, and behave like bare modes in either side of the strong coupling region. The TP and the cavity mode repel each other strongly away from their unique uncoupled mode energy at different values of $d_s$, which results to the anticrossing of the two modes when they approach each other. This anticrossing is the signature of strong coupling. Therefore, the minimum thickness of S layer that is responsible for strong coupling of modes is $d_s = 141$ nm.

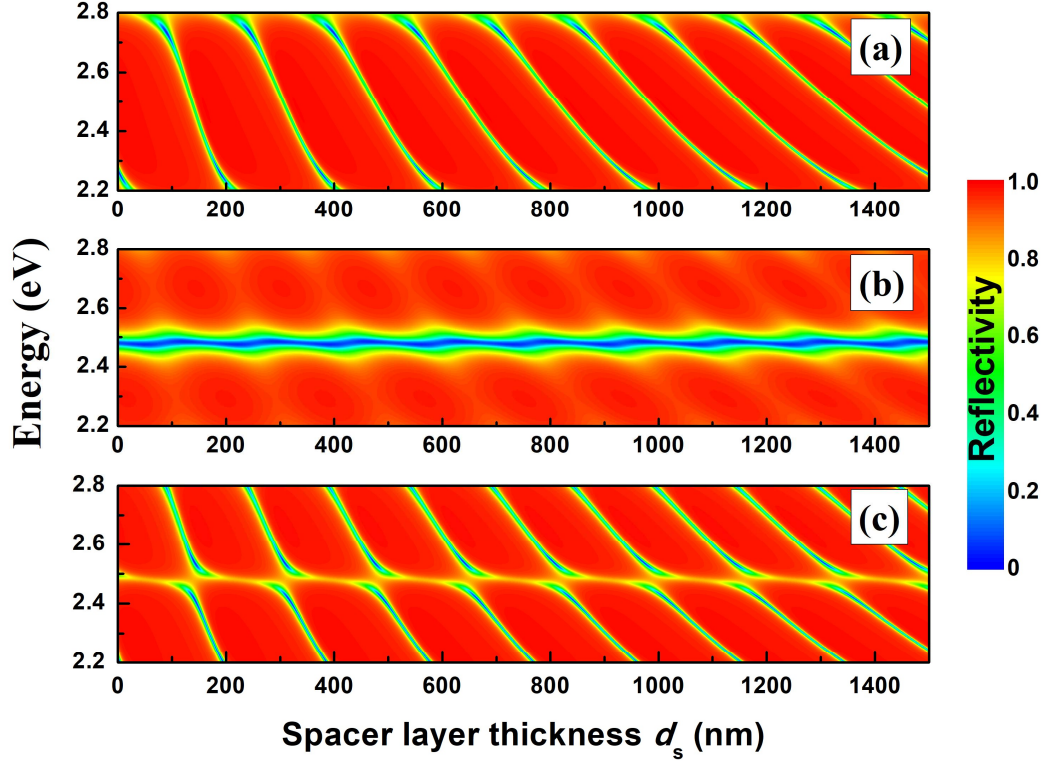

**Fig. S3**. Reflectivity contour as a function of incident light energy and S layer thickness $d_s$ for **(b)** MS(HL)$^{10}$, (b) (HL)$^5$C(HL)$^5$, and (c) MS(HL)$^5$C(HL)$^5$ photonic structure, respectively.

**Reflectivity of hybrid modes with varying number of periods (N):**

Fig. S4 shows that the cavity mode shifts to higher wavelength and its linewidth becomes wider with increasing value of $n_a$, while the TP mode remains unchanged. The refractive index of analyte is varied from 1.33 to 1.49 with an interval of $\Delta n=0.04$. With increasing N, the linewidth of the cavity mode gets sharper that increases its Q-factor. For lower N value, the modes are weakly localized as a result the reflectivity dip linewidth gets wider and the resonant TP mode changes with varying N. Therefore, such low N structures can no longer be useful for making sensors. For higher N value, the TP mode gets strongly localized as a result the corresponding dip reflectivity and wavelength values remain unchanged for higher value of N and it acts as a reference. Therefore, the hybrid structure can be used for self-referenced refractive index sensing.

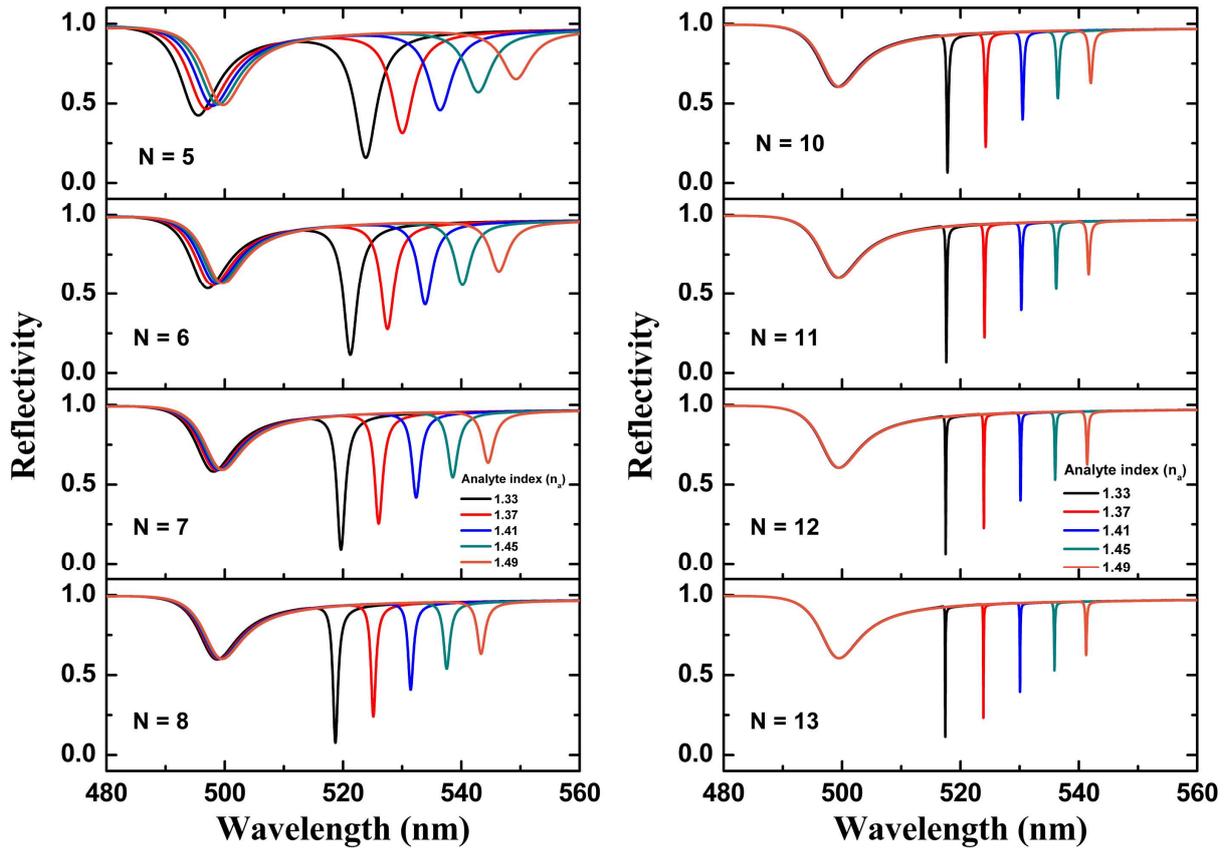

**Fig. S4.** Reflection spectra of the hybrid photonic structure with varying analytes refractive index ($n_a$) in the cavity layer as a function of number of period N.

**Reflectivity of Ag/SiO$_2$/(TiO$_2$/SiO$_2$)$^6$/C/(TiO$_2$/SiO$_2$)$^6$ with varying cavity thickness for two analyte indices $n_a$=1.33 and 1.49:**

In Fig. S5, the cavity layer C can be filled with different analytes. Here, we have only considered two analytes having extreme refractive index values $n_a$ =1.33 and 1.49. The figure shows that the coupling of modes depends on cavity thickness and there exists several values of cavity thickness for which the modes get strongly coupled and deviate from each other. Therefore, such thickness values should be avoided for sensing applications as both the coupled modes shifts their positions with varying analyte medium. For cavity thickness values other than strong coupling region, the reference mode $\lambda_R$ remains unchanged, while the analyte mode $\lambda_a$ shifts. Such cavity thickness values can be selected for self-referenced sensing applications. Here, 5-values of cavity thickness $d_c$=170 nm, 340 nm, 510 nm, 705 nm, and 892 nm, which lie in the weak coupling region, are selected for example purposes. The arrow indicated towards right side in the plot represents $\lambda_a$ above $\lambda_R$, while the arrow directed towards left side represents $\lambda_a$ below $\lambda_R$. One can clearly see the shifting of both lower and upper analyte mode with varying refractive index of analyte from 1.33 to 1.49 for the selected values of cavity thickness, which is explored for refractive index sensing application.

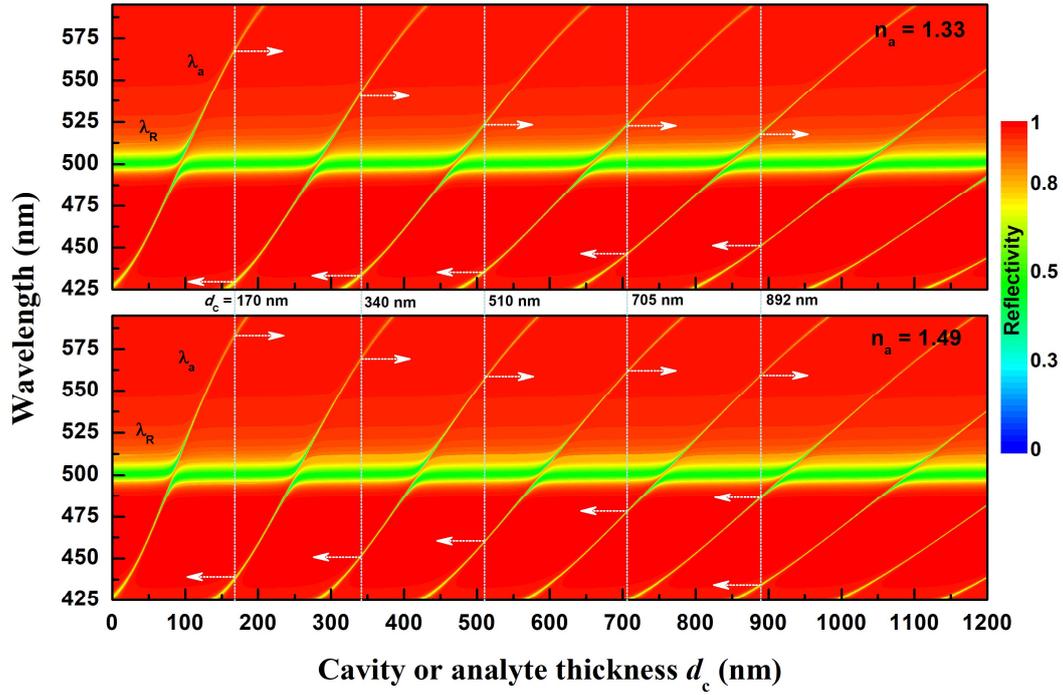

**Fig. S5.** Reflectivity contour as a function of wavelength and cavity or analyte thickness $d_c$ for two different values of refractive index of analyte: $n_a = 1.33$ and 1.49, respectively in a hybrid structure of MS(HL)$^N$C(HL)$^N$ with (HL)$^N$ = (TiO$_2$/SiO$_2$)$^6$.

**Reflectivity of two hybrid structures S1 and S2 for wide spectral range:**

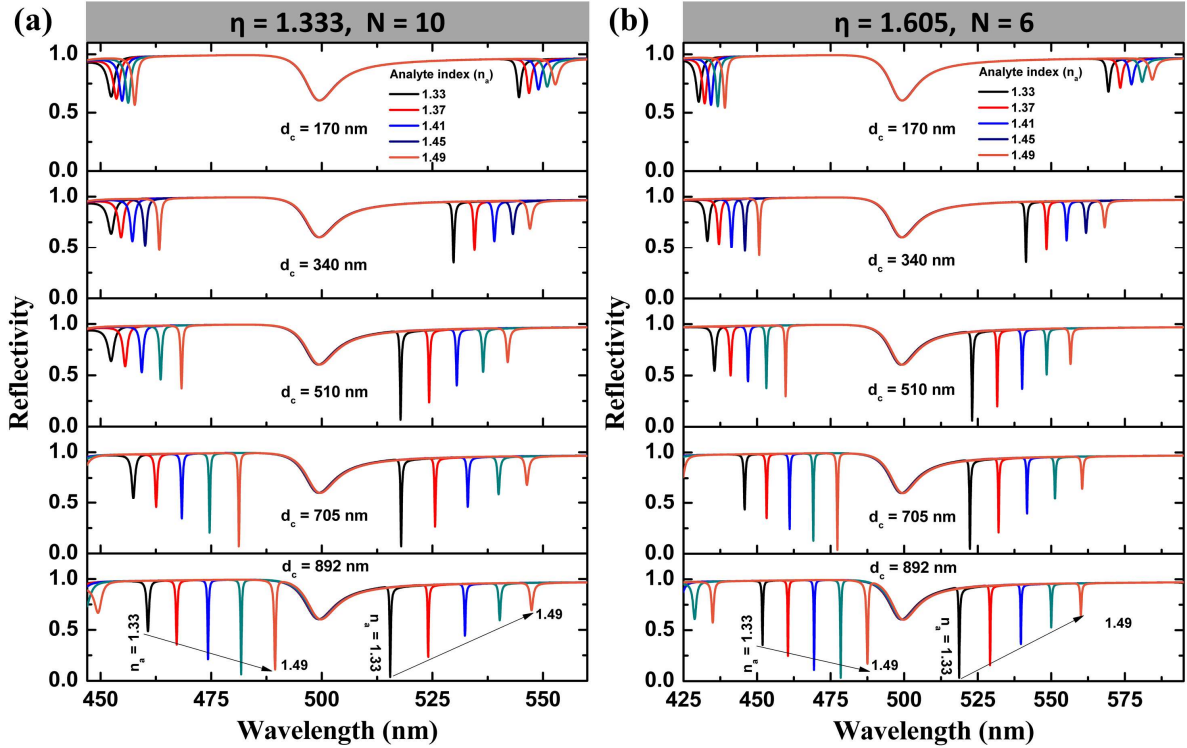

**Fig. S6.** Wider spectral reflectivity of hybrid structures S1 and S2 with varying refractive index of analyte filled in the cavity for different cavity thicknesses. Cavity mode either above or below TP mode can be used for sensing application.